\newcommand{\arxiv}{1} 

\documentclass[times,twocolumn,final]{elsarticle}

\usepackage{marvosym}
\usepackage{anyfontsize}
\usepackage{setspace}

\ifdefined\arxiv
    \usepackage{medima-arxiv}
    
    \newcommand{\orcid}[1]{\href{https://orcid.org/#1}{\includegraphics[scale=0.09]{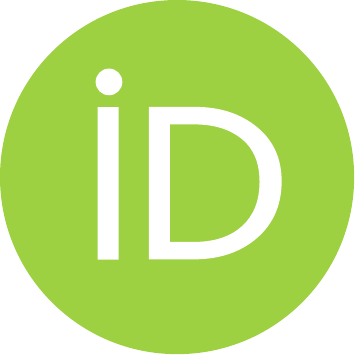}}}
    \newcommand{\mailto}[1]{\href{mailto:#1}\textsuperscript{\textsuperscript{\Letter,}} }
    \journal{}
    \date{}
    \newcommand{\hint}[1]{}
    \newcommand{\writer}[1]{~}
    \renewcommand{\snm}[1]{#1}
    \newcommand{\subtitle}[1]{\newline\normalfont\itshape-~#1~-}
    \newcommand{\rebuttal}[1]{#1}

\else
    \usepackage{medima}
    \newcommand{\writer}[1]{~}
    
    \newcommand{\orcid}[1]{}
    \newcommand{\mailto}[1]{}
    \newcommand{\hint}[1]{}
    \newcommand{\subtitle}[1]{{\normalfont-~}#1}
    \newcommand{\rebuttal}[1]{\color{blue} #1}
\fi

\usepackage{framed}
\usepackage{multirow}
\usepackage{multicol}
\usepackage{tabularx}
\usepackage{booktabs}
\usepackage{amssymb}
\usepackage{amsmath}
\usepackage{latexsym}
\usepackage[obeyspaces]{url}
\usepackage[dvipsnames]{xcolor}
\usepackage{rotating}
\usepackage{makecell}
\usepackage{pifont}
\usepackage{array}
\usepackage{etoolbox}
\usepackage{colortbl}
\usepackage{blindtext}
\usepackage{wrapfig}
\usepackage[colorlinks=true,citecolor=blue,urlcolor=blue]{hyperref}
\AtBeginDocument{%
  \hypersetup{
    citecolor=teal,
    linkcolor=blue,   
    urlcolor=Blue}}

\newcommand{\triplet}[1]{\textlangle{\textit{#1}}\textrangle{}}

\newcolumntype{P}[1]{>{\centering\arraybackslash}p{#1}}

\preto\tabular{\setcounter{magicrownumbers}{0}}
\newcounter{magicrownumbers}

\definecolor{newcolor}{rgb}{.8,.349,.1}
\journal{Medical Image Analysis}

\begin{document}

\verso{Chinedu I. Nwoye \textit{et~al.}}

\begin{frontmatter}

\title{{\bf \quad  CholecTriplet2022: Show me a tool and tell me the triplet \subtitle{an endoscopic vision challenge for surgical action triplet detection}}}%

\author[1]{ 
\orcid{0000-0003-4777-0857} Chinedu Innocent \snm{Nwoye} \mailto{nwoye.chinedu@gmail.com}} \corref{cor1} \cortext[cor1]{Corresponding author}

\author[1]{Tong \snm{Yu}}
\author[1]{Saurav \snm{Sharma}}
\author[1]{Aditya \snm{Murali}}
\author[1]{Deepak \snm{Alapatt}}
\author[1,2]{Armine \snm{Vardazaryan}}
\author[1,8]{Kun \snm{Yuan}}

\author[3]{Jonas \snm{Hajek}}
\author[3]{Wolfgang \snm{Reiter}}
\author[4]{Amine \snm{Yamlahi}}
\author[4]{Finn-Henri \snm{Smidt}}
\author[5]{Xiaoyang \snm{Zou}}
\author[5]{Guoyan \snm{Zheng}}
\author[6,60,600]{ Bruno \snm{Oliveira}}
\author[6,60,600]{Helena R. \snm{Torres}}
\author[7]{Satoshi \snm{Kondo}}
\author[70]{Satoshi \snm{Kasai}}
\author[8]{Felix \snm{Holm}}
\author[8]{Ege \snm{Özsoy}}
\author[9]{Shuangchun \snm{Gui}}
\author[9]{Han \snm{Li}}
\author[10]{Sista \snm{Raviteja}}
\author[10]{Rachana \snm{Sathish}}
\author[1100]{Pranav \snm{Poudel}}
\author[11,110]{Binod \snm{Bhattarai}}
\author[12]{Ziheng \snm{Wang}}
\author[12]{Guo \snm{Rui}}

\author[4,40]{Melanie \snm{Schellenberg}}
\author[6]{João L. \snm{Vilaça}}
\author[8]{Tobias \snm{Czempiel}}
\author[9]{Zhenkun \snm{Wang}}
\author[10]{Debdoot \snm{Sheet}}
\author[11000]{Shrawan Kumar \snm{Thapa}}
\author[12]{Max \snm{Berniker}}
\author[4,40]{Patrick \snm{Godau}} 
\author[6]{Pedro \snm{Morais}}
\author[11000]{Sudarshan \snm{Regmi}}
\author[4]{Thuy Nuong \snm{Tran}}
\author[600]{Jaime \snm{Fonseca}}
\author[4,40]{Jan-Hinrich \snm{Nölke}} 
\author[60]{Estevão \snm{Lima}}
\author[1100]{Eduard \snm{Vazquez}}
\author[4]{Lena \snm{Maier-Hein}} 
\author[8]{Nassir \snm{Navab}}  

\author[13]{Pietro \snm{Mascagni}}
\author[1,14,2]{Barbara \snm{Seeliger}}
\author[14,2]{Cristians \snm{Gonzalez}}
\author[14,2]{Didier \snm{Mutter}}

\author[1,2]{ Nicolas \snm{Padoy}}

\address[1]{ICube, University of Strasbourg, CNRS, France}
\address[3]{Riwolink GmbH, Germany}
\address[4]{Division of Intelligent Medical Systems (IMSY), German Cancer Research Center (DKFZ), Heidelberg, Germany}
\address[40]{National Center for Tumor Diseases (NCT), Heidelberg, Germany}
\address[5]{Institute of Medical Robotics, School of Biomedical Engineering, Shanghai Jiao Tong University, China}
\address[6]{2Ai School of Technology, IPCA, Barcelos, Portugal}
\address[60]{Life and Health Science Research Institute (ICVS), School of Medicine, University of Minho, Braga, Portugal}
\address[600]{Algoritimi Center, School of Engineering, University of Minho, Guimeraes, Portugal}
\address[7]{Muroran Institute of Technology, Japan}
\address[70]{Niigata University of Health and Welfare, Japan}
\address[8]{Technical University Munich, Germany}
\address[9]{Southern University of Science and Technology, China}
\address[10]{Indian Institute of Technology Kharagpur, India}
\address[11]{University College London, UK}
\address[110]{University of Aberdeen, UK}
\address[1100]{Redev Technology Ltd, UK}
\address[11000]{Nepal Applied Mathematics and Informatics Institute for research (NAAMII), Nepal}
\address[12]{Intuitive Surgical, USA}
\address[13]{Fondazione Policlinico Universitario Agostino Gemelli IRCCS, Rome, Italy}
\address[14]{University Hospital of Strasbourg, France}
\address[2]{IHU Strasbourg, France}

\received{: }
\finalform{:}
\accepted{:}
\availableonline{:}
\communicated{:}

\begin{abstract} 
Formalizing surgical activities as triplets of the used instruments, actions performed, and target anatomies is becoming a gold standard approach for surgical activity modeling. 
The benefit is that this formalization helps to obtain a more detailed understanding of tool-tissue interaction which can be used to develop better Artificial Intelligence assistance for image-guided surgery.
Earlier efforts and the CholecTriplet challenge introduced in 2021 have put together techniques aimed at recognizing these triplets from surgical footage.
Estimating also the spatial locations of the triplets would offer a more precise intraoperative context-aware decision support for computer-assisted intervention.
This paper presents the {\it CholecTriplet2022} challenge, which extends surgical action triplet modeling from recognition to detection. It includes weakly-supervised bounding box localization of every visible surgical instrument (or tool), as the key actors, and the modeling of each tool-activity in the form of \triplet{instrument, verb, target} triplet.
The paper describes a baseline method and 10 new deep learning algorithms presented at the challenge to solve the task.
It also provides thorough methodological comparisons of the methods, an in-depth analysis of the obtained results across multiple metrics, visual and procedural
challenges; their significance, and useful insights for future research directions and applications in surgery.
\end{abstract}

\begin{keyword}
\vspace{-0.3in}
\KWD Action detection \sep tool localization \sep fine-grained activity recognition \sep surgical action triplet \sep weak supervision \sep CholecT50 \sep computer-assisted surgery.
\end{keyword}

\end{frontmatter}


\section{Introduction }

\begin{figure*}[thbp]
    \centering
        \includegraphics[width=\textwidth]{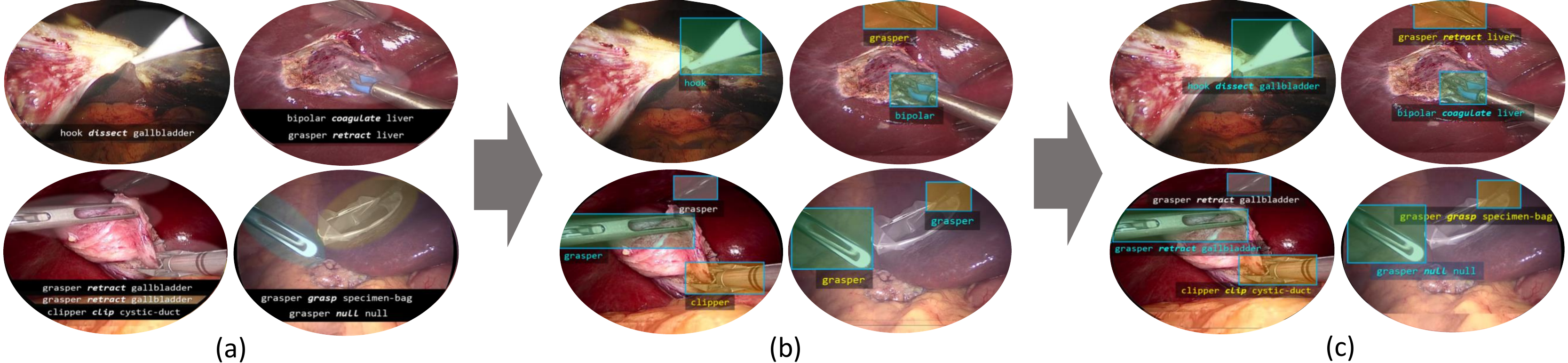} 
        \caption{Illustration of the 3 sub-tasks of the CholecTriplet2022 challenge on the CholecT50 dataset: (a) Triplet recognition: triplet binary labels, (b) Instrument localization: instrument binary + instrument spatial labels, and (c) Triplet detection: triplet binary + instrument spatial labels. Represented surgical action triplets are illustrated for different time points during laparoscopic cholecystectomy.}
    \label{fig:task}
\end{figure*}

\hint{give a general overview of the paper.}
Real-time video analysis will become an essential part of surgical intervention \citep{mascagni2022computer}. It can offer context-aware intraoperative decision support to surgeons using artificial intelligence (AI) computational models capable of extracting knowledge from live video data with reliability, accuracy, and speed \citep{vercauteren2019cai4cai,nwoye2021deep}.
The goal is to provide new intraoperative assistance thanks to the complementary analysis of the surgical workflow in terms of activity recognition \citep{twinanda2016endonet,lecuyer2020assisted}, tool (or instrument) detection and tracking \citep{jin2018tool,al2018monitoring,nwoye2019weakly}, tissue/tumor segmentation and classification \citep{luengo20212020,wang2022autolaparo,maqbool2020m2caiseg}, etc.
Despite the tremendous progress made in the research community, activity modeling, such as phase recognition \citep{czempiel2020tecno,gao2021transsvnet}, is still coarse-grained and does not contain the details needed for highly adaptive AI assistance in the operating room (OR). The fine-grained counterparts, such as action recognition \citep{khatibi2020proposing,wagner2021comparative}, leave out the details about the anatomy operated on.

The quest for a better and more comprehensive activity modeling led to the formalization of surgical activities as triplets
\citep{katic2014knowledge,nwoye2021deep} based on formal surgical ontology \citep{neumuth2006structured,katic2015lapontospm,gibaud2018toward}. These triplets take into account the intricate details of the operating instruments, the manipulated tissues or targets, and the verbs of actions that describe the interaction between them.
Earlier work on surgical action triplet recognition \citep{nwoye2020recognition,nwoye2021rendezvous,xi2022forest,li2022sirnet,cheng2022deep} including our previous CholecTriplet2021 challenge \citep{nwoye2022cholectriplet2021} concentrate primarily on the presence detection of these triplets from laparoscopic videos. 
Localizing also the spatial regions of action triplets across video frames would advance research on tool-tissue interaction understanding and help develop assistance systems delivering valuable AI feedback and automated warnings.
 
This paper presents \textbf{ CholecTriplet2022}\footnote{\url{https://cholectriplet2022.grand-challenge.org}}, an endoscopic vision challenge organized at MICCAI 2022 for the detection of surgical action triplets in laparoscopic videos. This is an extension of the previous edition of the international competition on triplet presence detection at MICCAI 2021 with \textbf{spatial localization} of the instruments performing the actions (illustrated in Fig \ref{fig:task}). 
The challenge was organized under the aegis of the Endoscopic Vision (EndoVis) grand-challenge \citep{aneeq-zia-2022-6390403} and presented at MICCAI 2022 in Singapore. The international challenge provided a scientific platform for the development of deep learning solutions for surgical action triplet detection in the OR. During the contest, registered participants were granted privileged access to a part of the CholecT50 dataset \citep{nwoye2021rendezvous} used as training data. The provided data labels are based on binary presence only, motivating the contestants to innovate by proposing alternative approaches, such as weak supervision, for modeling the localization aspect of the task. Meanwhile, some spatially annotated mini-samples of the dataset are provided for checks and validation. Useful code repositories, a validation system, a discussion forum, a submission server, support, and other helpful resources were also provided. A total of 11 teams participated in the challenge and presented several promising technologies
using AI to replicate a detailed understanding of tool-tissue interactions that experts in minimally invasive surgery have acquired through many years of training.

In addition to these contributions brought by the event itself, the post-challenge report offers contributions on its own.
Here, we study and present in this paper, a summary of the challenge activities, a theoretical description of all presented methods, an in-depth methodological analysis, and a method comparison. 
We benchmark the developed algorithms against their rivals and the baseline using the same criteria, and report the obtained results which are in the range of 18.8-35.0\% for triplet recognition, 0.3-41.9\% on instrument localization, and 0.08-4.49\% on triplet detection using average precision (AP) metrics at a threshold of 0.5 intersection over union (IoU), which is very strict for weakly supervised methods.
We also illustrate, using a rich selection of qualitative results, the behavior of proposed methods under various visual and procedural challenges associated with surgical video datasets. We highlight some strategic findings on the suitability of the observed computer vision techniques for surgical workflow activity modeling and discuss the significance and relevance of the presented results to the advancement of AI in surgery. 
We conclude with a survey polling the participants on their experiences and how to standardize and improve future events.

The rest of the paper is organized as follows: a review of related works in the next section helps to position our work in the research domain. This is followed by a summary of the challenge setup and activities. A detailed description, analysis, and comparison of the methods with the adapted evaluation protocols are presented in the subsequent sections.
Afterwards, the findings are analyzed and discussed in terms of their benefits and limitations. And the study concludes by discussing the future potential of the work done.
\section{Related work}
The CholecTriplet challenge relates to several research topics, for which we present the relevant literature in this section.

\subsection{Activity recognition}
Recognizing human activities from videos is a central task in computer vision, approached in different ways depending on the visual context - natural or medical images. Early methods from general computer vision employed architectures based only on 2D Convolutional neural networks (CNN): \cite{karpathy2014largescale} explored them in various pooling configurations, while \cite{simonyan2014twostream} used a pair of 2D CNN to simultaneously exploit RGB data and optical flow. More advanced forms of temporal modeling for activity recognition followed, starting with the long-term recurrent convolutional networks (LRCN) from \cite{donahue2015lrcn}: this architecture incorporated a long short time memory (LSTM) recurrent neural network as a model for temporal dependencies between frames. 3D CNN \citep{dutran2015c3d, carreira2017i3d, xie2018s3d, feichtenhofer2019slowfast} take a different approach by using spatio-temporal convolutions. Recent developments led to the video Transformers \citep{bertasius2021timesformer, liu2022videoswin}, built around spatio-temporal attention mechanisms.

In surgical computer vision, the most common activity recognition task is phase recognition. Several models were proposed for this task: \cite{twinanda2016endonet} used the same concept as the LRCN \citep{donahue2015lrcn}, combining a AlexNet model with a Hierarchical Hidden Markov Model (HHMM) on the Cholec80 dataset. Variations of the same architecture including deep residual models and LSTM-based approaches were employed in subsequent works on the same dataset \citep{funke2018temporal, yu2019learning, jin2018svrcnet, jin2021temporal, gao2021transsvnet}. Other types of temporal models were explored, such as temporal convolutional networks (TCN) by \cite{czempiel2020tecno}, as well as \cite{ramesh2021multi}, who, in addition to phase, carried out the finer-grained task of step recognition. Most recently, the Transformer models \citep{czempiel2021opera,gao2021trans} were explored for surgical activity recognition. 

Overall, surgical computer vision has mostly focused on coarse-grained, long-range activity recognition tasks, leading to models that are different from those found in natural computer vision.

\subsection{Action triplet: from recognition to detection}
Beyond coarse-grained activity recognition, activities that involve more than one component provide an exciting way to understand the complexity of an activity. Fine-grained activities are commonly denoted by a triplet of \triplet{\textit{subject, verb, object}} where subject denotes the actor, verb the type of activity, and object the end target of the activity. To this end, \cite{chao2015hico} analyzed a diverse set of human-object interaction instances. \cite{mallya2016learning} first extracted human and object region of interest features from a CNN and applied Multi-Instance Learning (MIL) technique for activity prediction. \cite{yuwei2018learning}, on the other hand, modeled individual and pairwise associations of detected objects along with graph-based label correlation in a multi-stream architecture. \cite{gkioxari2018detecting} utilized FasterRCNN to detect humans and objects in a multi-task learning setup. \cite{qi2018learning} adopted a graph structure with detected humans and objects as nodes in an iterative message-passing method. \cite{zou2021end} and \cite{tamura2021qpic} employed transformer-like architecture in an end-to-end learning approach with the help of matching loss and learnable queries. \cite{zhang2022efficient} proposed a two-stage unary and pairwise token-based transformer to analyze human-object interaction. 

Action triplets in the context of surgical computer vision was first formalized in \cite{neumuth2006structured} and subsequently adopted in \cite{katic2014knowledge,katic2015lapontospm} and \cite{forestier2015automatic}, as a valuable source of complementary information to help with the main task of surgical phase recognition.
Meanwhile, previous research has investigated multiple levels of granularity in surgical process modeling, ranging from high-level procedures \citep{sandberg2005deliberate,fischer2005ent} to more detailed steps, substeps, tasks, subtasks \citep{burgert2006linking,ramesh2021multi}, and motions \citep{lin2006towards,nomm2008recognition,ahmadi2009motif}. By utilizing these fine-grained labels, researchers can acquire a better understanding of the fundamental mechanisms involved in a surgical process. In fact, combining these fine-grained labels can yield a more comprehensive and accurate model of surgical processes \citep{lalys2014surgical}.

Action triplet recognition as a primary task was first introduced in \cite{nwoye2020recognition} where they designed unary branches to predict triplet components: instrument, verb, and target, and a learnable 3D interaction space to model an association between the components. Later, \cite{nwoye2021rendezvous} introduced an improved transformer-style model with an attention-based mechanism to further enhance triplet components association towards triplet recognition. \cite{xu2021learning} employed a transformer model along with adversarial learning to generate captions, akin to triplets, depicting semantic relationships between components involved in a surgical scene. \cite{lin2022instrument} assigned instrument and target bounding boxes to triplet information and utilized a spatio-temporal graph for instrument-target interaction detection in cataract surgery.

\subsection{Datasets: from recognition to detection}
Action datasets offer a large choice of tasks to be learned and performed by algorithms, with variations in granularity as well as the nature of the proposed task. From general computer vision, one early example of action classification datasets is PASCAL VOC \citep{everingham20052005}, with 10 action classes and static images only. To better capture actions, video datasets were released: UCF-101 \citep{soomro2012ucf101}, followed by Charades \citep{sigurdsson2016charades} and Kinetics \citep{carreira2017i3d}. Another solution for capturing more information on actions is to add spatial detection: V-COCO \citep{gupta2015vcoco} extended the original COCO dataset \citep{lin2014coco} with bounding boxes around interacting elements. HICO-DET \citep{chao2015hico}, centered around human-object interaction, offered similar annotations. Video datasets with spatial action detection, such as UCFSports \citep{soomro2014ucfsports} or AVA \citep{gu2018ava} exist as well.

Action recognition in the surgical domain mainly comes in the form of surgical phase classification, as proposed by the Cholec80 dataset \citep{twinanda2016endonet} dividing cholecystectomy into 7 phases. The CATARACTS dataset \citep{al2019cataracts} contained similar phase annotations for cataract surgery. Finer-grained descriptions of activities were featured in the Bypass40 dataset \citep{ramesh2021multi} via surgical steps. Additional information can also come in the form of spatial detections: \cite{vardazaryan2018tool, nwoye2019weakly} used an extension of Cholec80 with instrument bounding boxes on the test set, for the evaluation of weakly supervised object detectors; however, those only located actions in an indirect manner. Actual action localization was offered by the SARAS-ESAD dataset \citep{vivek2021thesaras,lin2022instrument}, with bounding boxes pointing to action verbs being performed.
Other closely related dataset such as JIGSAWS \citep{gao2014jhu} provides data labels of different granularity levels for surgical skill assessment, motion analysis, and automation.
The introduction of CholecT40 dataset \citep{nwoye2020recognition} and its successor CholecT50 \citep{nwoye2021rendezvous} expands upon these efforts by providing direct labeling of surgical activities in the form of \triplet{instrument, verb, target} triplets for surgical action triplet recognition.

Overall, action datasets have evolved towards more detailed and complex tasks, gradually adding more information in the form of interaction labels or bounding boxes. Our work continues in this direction by proposing a weakly supervised action localization task on the CholecT50 dataset \citep{nwoye2021rendezvous}.

\subsection{Benchmark challenge: from recognition to detection}
International challenges have become a de facto standard for benchmarking image analysis algorithms \citep{wiesenfarth2021methods}. 
They have a significant impact on the research community, oftentimes, encouraging a surge in new research directions owing to their characteristic creation of shared datasets, as well as incentives for best-performing algorithms.
And more significantly, they offer a more reliable result benchmarking by withholding their test data from the public domain and comparing rival algorithms using the same criteria. 
The PASCAL VOC challenge \citep{everingham20052005} and the ImageNet Large Scale Visual Recognition Challenge (ILSVRC) \citep{russakovsky2015imagenet} are among the earliest image analysis challenges featuring deep learning methods albeit focusing on simple classification.
Subsequent challenges advance the recognition task to object detection \citep{lin2014coco,everingham2010pascal}, single and multi-object tracking \citep{dendorfer2020mot20,chen2021visdrone,VOTTPAMI}, segmentation \citep{Cordts2016Cityscapes,voigtlaender2019mots}, etc., spearheading the development of state-of-the-art models.
Repeating some challenges over the years comes with either or both increasing the size of datasets and task difficulty.

The first MICCAI grand challenge organized in 2007 in Brisbane brought similar benefits to biomedical image analysis. 
Notable in the surgical data science domain is the M2CAI 2016 challenge\footnote{\url{http://camma.u-strasbg.fr/m2cai2016/}} featuring both surgical workflow analysis and tool presence detection for laparoscopic cholecystectomy leading to the creation of widely used datasets in the field: m2cai16-workflow \citep{stauder2016tum,twinanda2016endonet}, m2cai16-tool and Cholec80 \citep{twinanda2016endonet}.
Other datasets such as BraTS \citep{menze2014multimodal}, CATARACTS \citep{al2019cataracts}, ROBUST-MIS \citep{RO2021101920}, SurgVisDom \citep{zia2021surgical}, MISAW \citep{huaulme2021micro}, HeiChole \citep{wagner2021comparative}, CholecT50 \citep{nwoye2021rendezvous}, SARAS-ESAD \citep{vivek2021thesaras}, Robotic Instrument Segmentation \citep{allan20192017}, CaDIS \citep{grammatikopoulou2019cadis,luengo20212020}, etc, are all products of biomedical challenges.

In 2021, we introduce the first endoscopic vision challenge focusing on the recognition of surgical activities in the form of triplets \citep{nwoye2022cholectriplet2021}.
A rerun of this challenge in the current edition is with two notable advances: (1) the increment of the task difficulty from presence to spatial detection, which also increases the task usefulness, and (2) the addition of spatial bounding box labels to challenge evaluation data thereby enriching the dataset.
\section{Challenge description}

\begin{figure*}[tp]
    \centering
    \includegraphics[width=\linewidth]{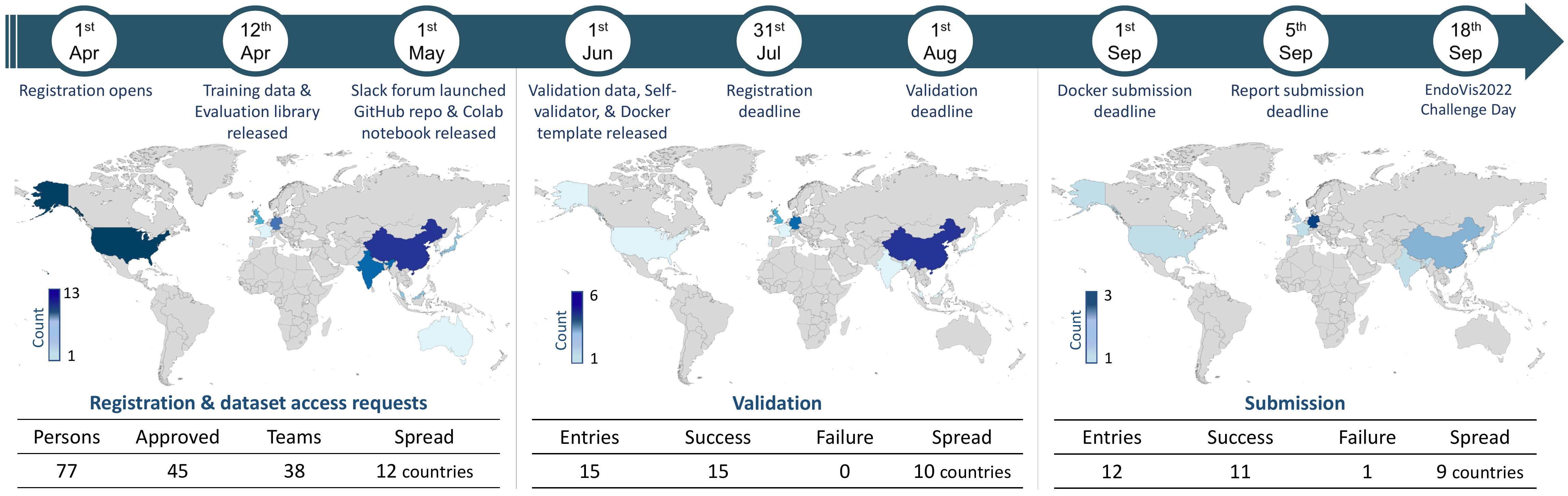}
    \caption{CholecTriplet2022 challenge timeline of activities and participation statistics.}
    \label{fig:timeline}
\end{figure*} 

CholecTriplet is an endoscopic vision challenge that orchestrates research, development, and evaluation of AI methods for the automatic analysis of surgical video activities syntactically as \triplet{instrument, verb, target} triplets; also known as surgical action triplet \citep{nwoye2020recognition}.
Its inception in 2021 under the code-name of {\it CholecTriplet2021} schemes to the presence detection (or simply, \textit{recognition}) of action triplets directly from surgical videos.
The current edition, {\it CholecTriplet2022}, organized by an 8-person committee from the Research Group CAMMA, advances the previous edition to also include the bounding box \textit{localization} of regions of action triplet's likelihood in laparoscopic videos.
We describe in detail the organization and execution of the CholecTriplet2022 challenge in this section.

\subsection{Task}
The primary goal is to develop machine learning methods for the \textit{detection} of surgical action triplets directly from surgical videos. The term ``detection'' in this context translates to joint recognition and localization tasks. This added to the complexity of surgical action triplet modeling coupled with the required simultaneous identification of the correct components (\textit{vis-à-vis}: instruments, verbs, targets), and resolving their association in the case of multi-instance triplets per frame.

The task is categorized into 3 sub-tasks (as shown in Fig \ref{fig:task}):
\begin{enumerate}
    \item {\bf Triplet recognition}: identification of the correct triplets, including their components, in every video frame,
    \item {\bf Spatial box localization}: estimation of the bounding box location of the principal actor (the instrument's tip) in every recognized triplet.
    \item {\bf Box-triplet association}: pairing of every localized instrument's bounding boxes to their corresponding triplets.
\end{enumerate}
During the challenge, the three sub-tasks are jointly treated as a single task: a submission, comprising either a single model or collaborating multiple models, must produce the three outputs in a single docker run to be considered complete.

\subsection{Challenge design}
The CholecTriplet 2022 challenge is designed following the BIAS Reporting Guideline \citep{maier2020bias} for enhanced quality and transparency of biomedical research. The structured design is submitted as part of the Endoscopic Vision (EndoVis) grand challenge \citep{aneeq-zia-2022-6390403} in December 2021. The proposal was approved after two rounds of MICCAI review followed by a call for participation circulated online and offline.

The challenge was officially launched on April 1, 2022, and run through a 6-months window as shown in Fig. \ref{fig:timeline}. 
This period is characterized by several activities such as the release of training data, customized metrics library, slack communication channel, a snippet of ``getting started" code, GitHub repositories, etc., to guide and support the participants' method development. 
The participating teams develop their novel methods, fine-tune a state-of-the-art method, or improve on existing solutions during this period.
The challenge timeline also involves a validation phase, harnessed by the use of a self-validation system, validation data samples, a Docker template, and guidelines provided to facilitate method submission.
The whole process is concluded with the presentation of the method, results, and award winners at MICCAI 2022 conference in Singapore on Sept. 18, 2022.

\subsection{Dataset and spatial annotation}
The challenge experiments are conducted on CholecT50 \citep{nwoye2021rendezvous}, the largest endoscopic video dataset for surgical action triplet recognition. 
The dataset consists of 50 video footages of laparoscopic cholecystectomy that has been annotated with 100 distinct categories of surgical action triplets. 
At 1 frame per second (fps), a total of 100.9K frames of the dataset has been annotated with ${\sim}{151K}$ triplet instances formed from 6 instrument, 10 verb, and 15 target categories.
We follow the official data splits \citep{nwoye2022data} for challenge purposes (\textit{aka} CholecT50-challenge version) which ensures that the test set is drawn from only the 5 videos that are not in the public domain for fairness in the competition. 
The rest of the 45 videos, \textit{aka} {CholecT45}, are released to the participants for training their models.
The CholecT45 provides only binary presence labels for the triplets, its individual components (instruments, verbs, targets), and the phase labels. 
The phase labels are provided as the coarse-grained level definition of the triplet activities and could be leveraged by deep learning models to improve triplet recognition.
Without spatial labels in the training data, the challenge allows for the modeling of the instrument's localization by weak supervision, thus, alleviating the cost and tedious annotation effort. 
While pretraining/transfer learning on third-party spatially annotated datasets is allowed, the downstream task involves learning the instrument localization and triplet-box association from imperfect annotations.

\begin{figure}[!tp]
    \centering    \includegraphics[width=.97\linewidth]{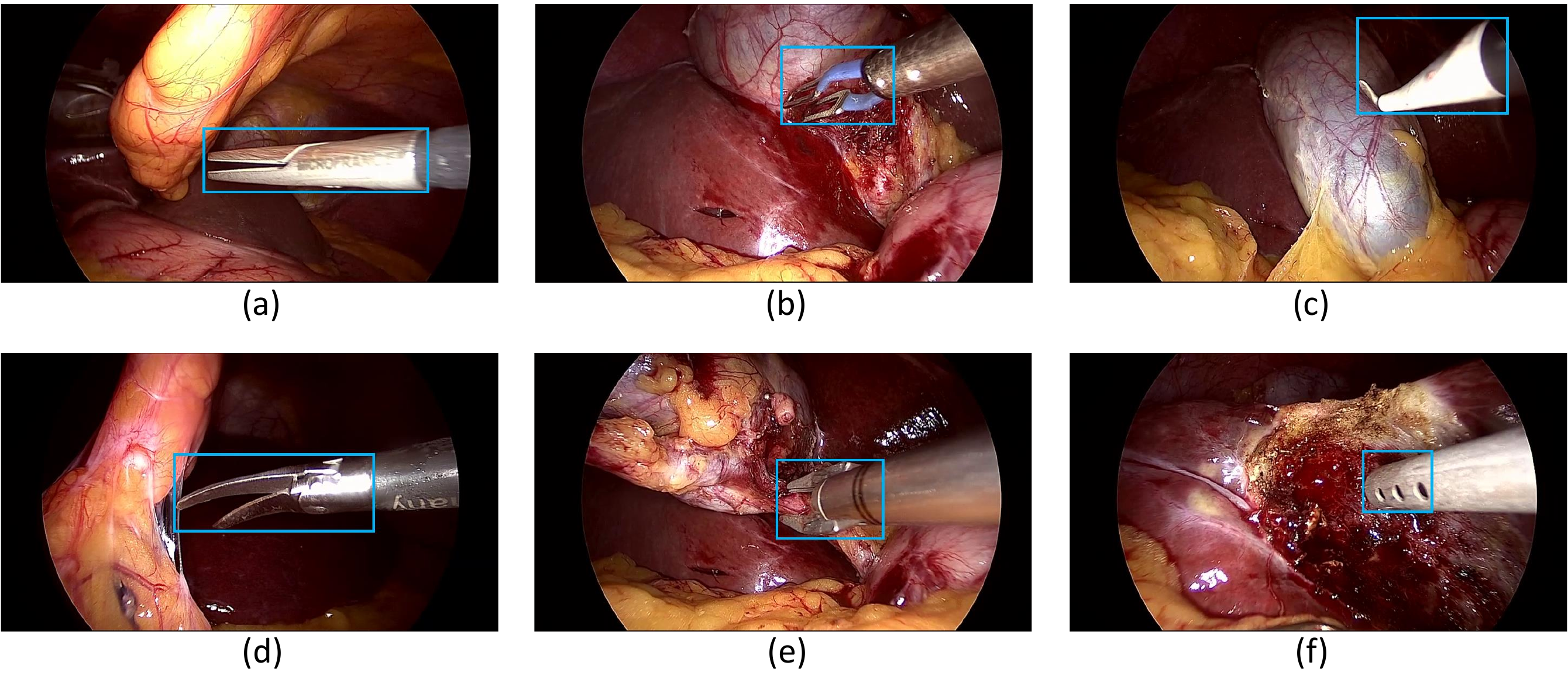} 
    \caption{Bounding box placements on the tip of the 6 instruments (a) grasper (b) bipolar - tip separated by a ribbon (c) hook (d) scissors (e) clipper - tip is separated from handle by two lines (f) irrigator - tip consists only the region with nozzles.}
    \label{fig:anns}
\end{figure}

For evaluation, we annotate the test set with bounding boxes over the instruments using the IHU MOSaiC web-based video annotation tool for surgical datasets. 
Conventionally, an instrument can be divided into two sub-parts: \textit{effector}, which is the instrument tip responsible for primary action on anatomy, and \textit{shaft} which connects the instrument tip with the instrument handle on its other end. As an annotation protocol, we consider only the effector part of the instrument and draw bounding boxes for instrument tips: (\textit{cold}) grasper, bipolar (\textit{grasper}), (\textit{monopolar}) hook, (\textit{monopolar}) scissors, clipper (short for \textit{clip applier}), and irrigator (short for \textit{suction/irrigation device}). 

The instruments, namely grasper, bipolar, hook, clipper, and scissors, have distinct boundaries that separate their tips from the shaft. The bipolar instrument's tip is separated by a ribbon, whereas the clipper's tip is distinguished by two lines. As for the irrigator, which lacks a clear demarcation between the tip and the shaft, a bounding box is drawn to tightly contain all visible nozzles. Figure~\ref{fig:anns} shows the bounding box annotation format for each category of instrument.

To ensure the accuracy of our annotations, we enlist the help of 5 computer scientists who follow a guideline verified by surgical experts. The annotation process consists of four stages: bounding box annotation, bounding box correction, box-triplet pairing, and box-triplet correction. The triplet labels had already been annotated by two surgeons in a previous study~\cite{nwoye2021rendezvous}. The annotators pair the bounding boxes with their corresponding triplet labels. As an aid to the correction task, we develop a Python-based visualization tool that allows for the overlay of all the labels on their respective images to ensure the high quality of our annotations. Disagreements are resolved with the aid of the visualization tool. The final correction is done by two annotators, and the resulting annotations are stored in JSON format, which includes both triplet binary presence labels and bounding box details.

\begin{table}[!tp]
    \centering
    \caption{CholecT50 dataset configuration used in the challenge}
    \label{tab:data-stat}
    \setlength{\tabcolsep}{15pt}
    \resizebox{0.85\columnwidth}{!}{%
    \begin{tabular}{@{}lrcr@{}}
        \toprule
         &  Training & Validation & Testing \\  
        \midrule
        \# full videos & 45 & - & 5\\
        \# short clips & - & 5 & - \\
        \# frames &  90.5K &  1.1K & 10.4K \\ \midrule
        \# triplet classes & 100 & 100 & 100 \\
        \# instrument classes & 7 & 7 & 7 \\
        \# verb classes & 10 & 10 & 10 \\
        \# target classes & 15 & 15 & 15 \\
        \# phase classes & 7 & 7 & 7 \\ \midrule
        \# triplet labels & 137.9K & 1.3K & 13.0K \\
        \# instrument labels & 137.9K & 1.3K & 13.0K \\
        \# verb labels & 137.9K & 1.3K & 13.0K \\
        \# target labels & 137.9K & 1.3K & 13.0K \\
        \# phase labels & 90.5K & 1.1K & 10.4K \\
        \# bounding boxes & - & 1.3K & 13.0K \\
        \bottomrule 
        \multicolumn{4}{l}{\makecell[l]{\scriptsize{\makecell[l]{Training set = CholecT45. Validation set = subset(CholecT45 ~$\cap$~ m2cai16-tool-location).}}
        }
        }
    \end{tabular}
    }
\end{table}

To give participants an insight into the testing labels, a mock-up validation set is generated. This involves 5 short video clips with triplet binary presence labels, instrument's bounding box labels, and box-triplet matching labels. The validation set's spatial annotation is outsourced from the overlapping videos of the m2cai16-tool-location dataset \citep{jin2018tool} and merged with the CholecT45 binary labels.
The statistics of the entire dataset as used in the challenge are provided in Table \ref{tab:data-stat}.

To access the training and validation datasets, participants first register on the challenge website and sign a non-disclosure contract on the usage of the dataset. Afterward, participants are provided with a download link to the online repository containing the dataset.

\subsection{Validation system} \label{sec:validator}
\begin{figure}[!tp]
    \centering
    \includegraphics[width=0.95\linewidth]{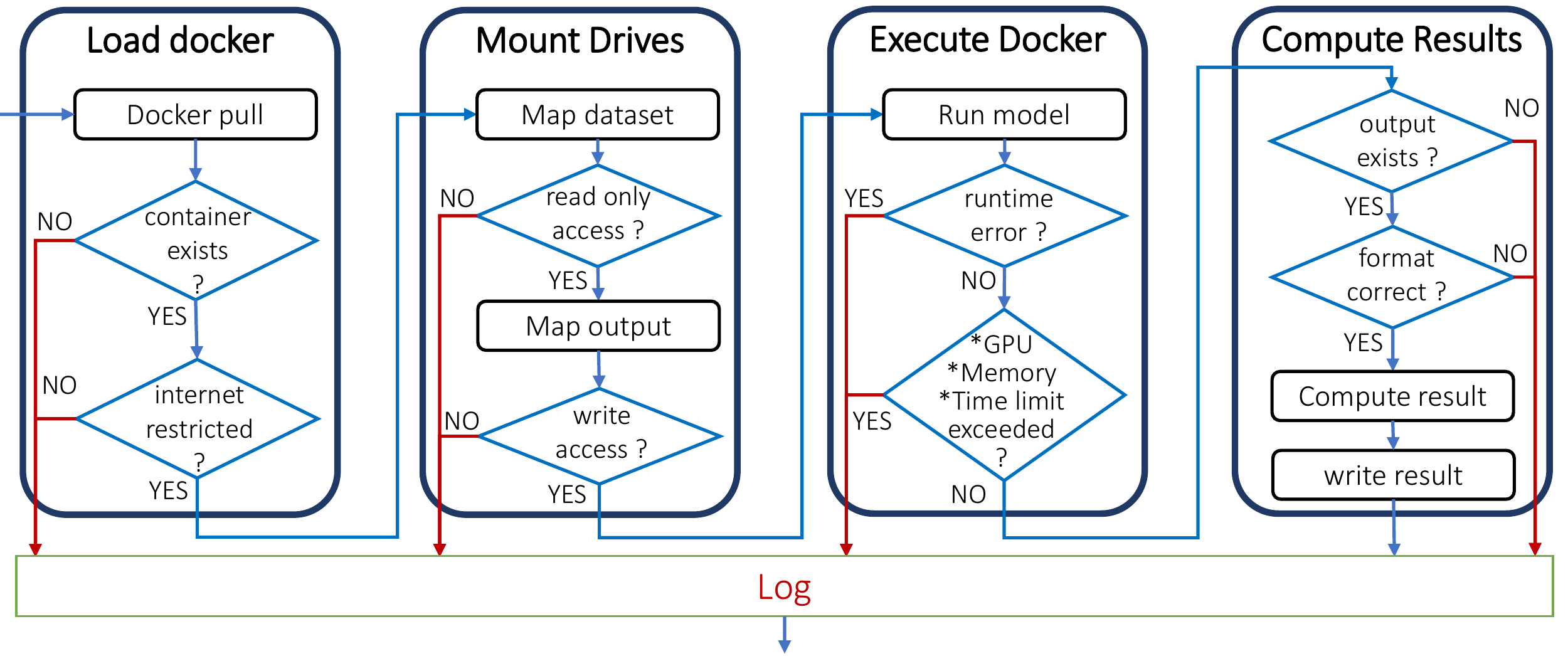}
    \caption{Flowchart of the self-validation system used in the challenge}
    \label{fig:validator}
\end{figure}

In the validation step, participants were tasked with testing their docker containers using a validation script, to ensure that they properly ingested input data as well as adhered to the proper specifications for their model outputs.
This validation script, implemented in Python, ensured that the participants' models respected the specified compute constraints, loaded and output data properly, and ran within the specified evaluation time and resource limits. The pipeline of the validation system is illustrated in Figure \ref{fig:validator}.

To aid participants, we split the validation phase into two parts: a local phase and a host phase.
During the local validation phase, we gave the participants access to a properly formatted validation dataset which they could use as a reference to develop their models and docker container.
Then, during the host testing phase, the participants were tasked with using the provided validation script to evaluate their finalized docker containers. 
This validator outputted a success report with performance on each test; participants were required to submit a successful validation report to the challenge submission portal before submitting their final docker container.
The whole self-validation concept is to mimic the actual evaluation server but on the participant's personal computer (PC).
The validation phase was highly effective, as 91.7\% of the finally submitted docker containers ran directly during the submission phase as can be seen in Fig. {\ref{fig:timeline}}; this is in contrast to the 2021 edition of the challenge, wherein 105 docker image submissions were required to successfully validate 24 teams.
\begin{table*}[!t]
    \centering
    \caption{~A cross-view of the methods presented by the teams at MICCAI EndoVis CholecTriplet 2022 challenge.}
    \label{tab:teams}
    \setlength{\tabcolsep}{6pt}
    \resizebox{\textwidth}{!}{%
    \begin{tabular}{@{}rllll@{}}
    \toprule & {\bf Network} & {\bf Method}  & {\bf Team } & {\bf Affiliation(s)} \\\toprule
 
    1 & AtomTKD & \makecell[tl]{A multiple atomic tasks knowledge distillation framework\\for triplet recognition and detection}  & SHUANGCHUN & Southern University of Science and Technology, China \\\midrule 
     
    2 & DATUM & \makecell[tl]{Detection of action triplets using multi-graph networks}  & KLIV-IITKGP & Indian Institute of Technology Kharagpur, India \\\midrule
     
    3 & Distilled-Swin-YOLO & \makecell[tl]{Self-distilled swin transformer ensemble } & SDS-HD & German Cancer Research Center (DKFZ), Germany \\\midrule
     
    4 & DualMFFNet & \makecell[tl]{Weakly-supervised surgical action triplet detection with\\dual multiplicative feature fusion networks}  & SK & \makecell[tl]{Muroran Institute of Technology, Japan\\Niigata University of Health and Welfare, Japan}\\\midrule
     
    5 &  EndoSurgTRD & \makecell[tl]{Multi-task spatial-temporal triplet recognition with\\ weakly-supervised tool detection}  & INTUITIVE-CORTEX-ML & Intuitive Surgical, USA \\\midrule
     
    6 & IF-Net & \makecell[tl]{Instrument first: multi-instance instrument detection \\and instance-wise triplet classification}  & CAMP & Technical University Munich, Germany \\\midrule
     
    7 & MTTT & \makecell[tl]{Multi-task triplet transformer for end-to-end surgical\\action triplet recognition}  & CITI & 
    \makecell[tl]{Institute of Medical Robotics, School of Biomedical\\ Engineering, Shanghai Jiao Tong University, China}\\\midrule
     
    8 & ResNet-CAM-YOLOv5 & \makecell[tl]{Combining ResNet class activation mapping with YOLOv5\\for instrument detection and triplet recognition} & WINTEGRAL & Riwolink GmbH, Germany \\\midrule
     
    9 & RDV-Det & \makecell[tl]{Rendezvous-det: A baseline extension of the rendezvous\\network for surgical action triplet detection}  & CAMMA & Universite de Strasbourg, France \\\midrule
     
    10 & SurgNet & \makecell[tl]{Surgical triplet recognition and detection using an ensemble\\of multi-task recurrent convolutional neural networks} & 2AI-ICVS & Applied Artificial Intelligence Laboratory, Portugal \\\midrule
     
    11 & URN-Net & \makecell[tl]{Mullti-task transformer with learnable orthogonal queries\\for triplet classification}  & URN & University College London, UK \\
        \bottomrule
    \end{tabular}
    }
\end{table*}
We attribute this difference to the inclusion of the local validation phase, which accelerated docker container development, as well as having provided the validation script directly to participants, which increased evaluation transparency.

\subsection{Submission protocol}
\hint{1. Docker design, hub, etc.
2. Submission format and portal,
3. Final evaluation runs (counts: success rate)}
Method submission is based on a Docker image uploaded via a dedicated challenge DockerHub. Private repositories are created for the teams with access granted by tokens valid until 5th September 2022. 
Smooth submission is enabled with a guideline published on the challenge GitHub page providing a Docker template with basic libraries, such as ivtmetrics, scikit-image, etc., required to run the model and evaluate their outputs. The template follows the same format used in the validation phase allowing participants to locally validate the Docker containers before pushing them to the submission portal. Submission update is possible only before the deadline.
Along with the method Docker, a success self-validation log file of the final Docker, a draft summary report, and a PowerPoint/video presentation of the proposed method are required to complete a team submission.
Afterward, all uploaded Docker containers are evaluated on the private test data (CholecT5), and results are collated for task benchmarking and team ranking.
Only a single (last) submission is evaluated for each team.

\subsection{Participation statistics}
During the call for participants, a total of 32 teams registered to participate in the challenge. Of these teams, a total of 11 teams progressed beyond the validation phase; as detailed in Fig \ref{fig:timeline}, these teams were drawn from 9 different countries and 3 continents.
This final participation was just over half of the CholecTriplet 2021 challenge~\citep{nwoye2022cholectriplet2021}; we also observed a drop in the relative ratio between the final participants and total registrants (45.5\% in 2021; 34.4\% in 2022).
We identified 2 likely reasons for the drop in each of these statistics: (1) the added difficulty of surgical triplet localization, which could have deterred registered teams from progressing to the validation phase, and (2) the use of almost the same dataset as in the 2021 edition, which may have reduced the perceived novelty of the challenge.
The participating teams and their method are presented in Table~\ref{tab:teams}.

\subsection{Awards}
There are several IHU-sponsored monetary and certificate awards for the winners and runners-up of each task category. 
Additionally, there is an NVIDIA-sponsored GPU award for the winner of the main task (category 3).
These awards are targeted at motivating and accelerating the research of the awardees even further.
\section{Methodology}
A summary of each participating team's proposed method is provided in Table~\ref{tab:teams}; in the following subsections, we elaborate on these summaries, providing detailed descriptions of each method and a comparative analysis.

\subsection{Analytical description of the conceptual frameworks}

\subsubsection{Rendezvous-Det (RDV-Det)}

\begin{figure}[!t]
    \centering        
    \includegraphics[width=.8\linewidth]{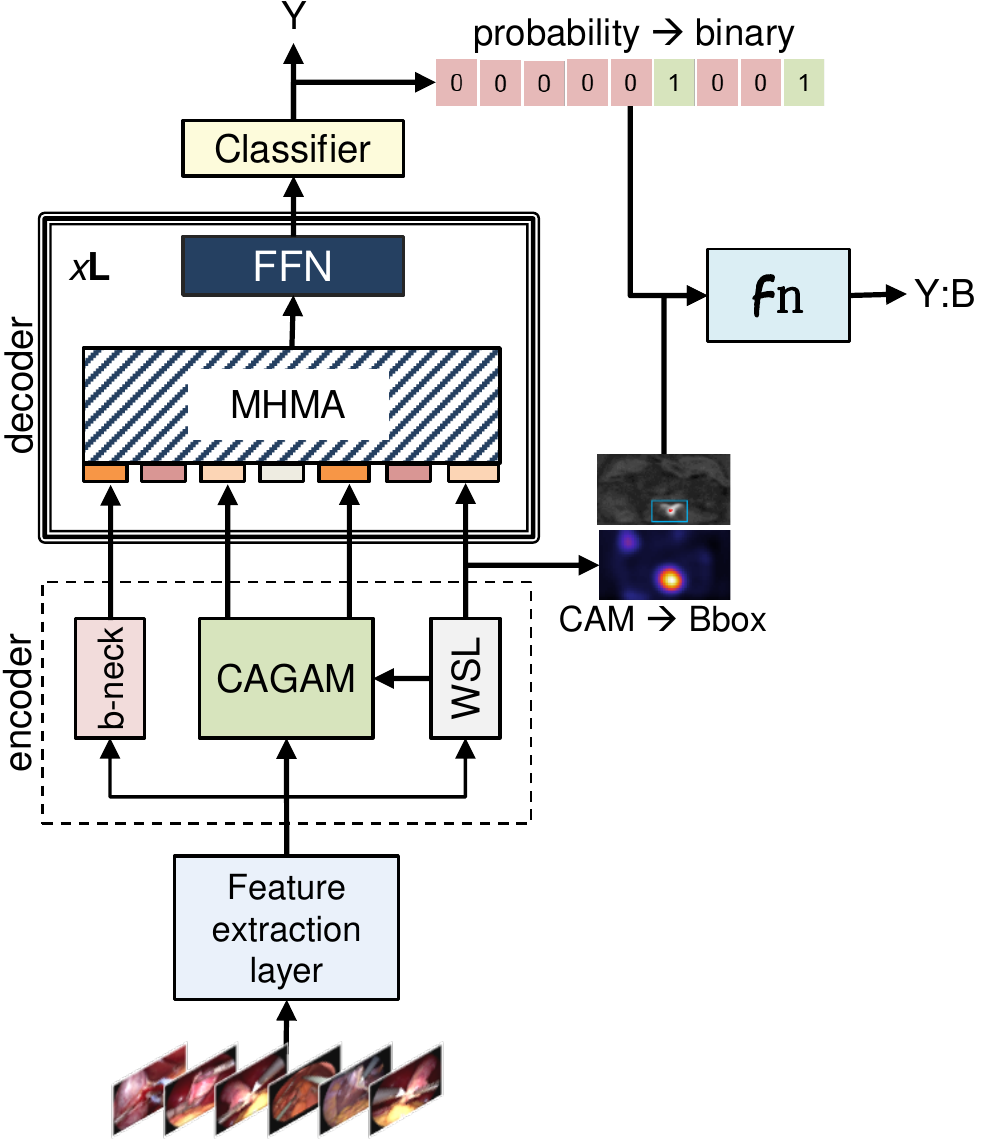} 
        \caption{
        Architecture and data flow of the RDV-det model. It features an attention mechanism, multi-task learning, class activation modeling, weak supervision, and linear assignment.        
        }
    \label{fig:rdv-det}
\end{figure}

This is proposed by the challenge organizers as a proof-of-concept and as a baseline to the competing methods. 
The RDV-det, shown in Fig. \ref{fig:rdv-det}, is a detection version of the Rendezvous \citep{nwoye2021rendezvous} model popular for surgical action triplet recognition. Notably, the Transformer-inspired RDV is conceptually made up of 3 major components: (1) a weakly supervised localization (WSL) layer for estimating the positions of surgical instruments in video frames, (2) a class activation guided attention module (CAGAM) for learning the verbs and targets leveraging an attention built from instruments activations, and (3) a multi-head of mixed attention (MHMA) for learning to resolve the instrument-verb-target relationships forming triplets. The MHMA is terminated by a linear classifier producing the triplet log probability scores, {\bf log Y}. 
Without loss of generality, we ignore other modules of the RDV such as: the feature extraction base model, bottleneck layer, projection layer, etc., that are not of paramount interest in the discussion of RDV-det.

To advance this model for detection, the instruments are first localized via the WSL layer. Here, the interest is in the last 6-channel convolution layer designed to learn the location of each of the six distinct instrument categories in the CholecT50 dataset in the form of class activation maps (CAM). As a post-process, we extract bounding box coordinates, {\bf B}, for every positive activation in the CAMs. Non-maximum suppression is applied to remove noisy labels.
Afterward, triplet binary presence scores, {\bf Y}, are obtained by thresholding the class probability vector of {\bf log Y}.
Finally, a heuristic-based data assignment module, {\bf fn}, is employed to associate every extracted bounding box, {\bf b $\in$ B}, to the corresponding triplet instance {\bf y $\in$ Y} within a given frame:
\begin{equation}
    y:b  \Leftarrow {\bf fn}(y_i,b_j, \phi) ~~~ \forall~ i=1..|{\bf Y}|, ~ j=1..|{\bf B}|,
\end{equation}
leveraging the instrument categories and triplet's detection confidence scores as the matching features, $\phi$, where $|{\bf Y}|$ and $|{\bf B}|$ are the cardinalities of Y and B respectively.

\subsubsection{AtomTKD}
AtomTKD is a teacher-student approach to triplet recognition in which the student model, S-TRNet, is trained to directly predict triplets while being additionally supervised by a teacher model, T-AtomNet, trained to predict each individual (atomic) triplet component (instrument, verb, and target) in a multi-task fashion.
Both S-TRNet and T-AtomNet consist of an ImageNet-pretrained ResNet18 backbone followed by a four-stage decoder consisting of a Transformer and convolutional layers \citep{dai2022ms}; the respective loss functions are applied to the outputs of each stage, resulting in a hierarchical model structure.
For the student model S-TRNet, the architecture is modified to use causal dilated convolutions, greatly reducing the number of model parameters.
Both S-TRNet and T-AtomNet are trained with binary cross-entropy loss functions: on triplets for S-TRNet, and separately on instrument, verb, and target for T-AtomNet.
Meanwhile, for the teacher-driven supervision signal, the predicted triplet logits from S-TRNet are filtered to obtain separate instrument, verb, and target prediction logits (e.g. aggregating triplet logits across all verbs and targets yields instrument logits), which are supervised by the respective logits predicted by T-AtomNet using MSE loss functions.

For triplet localization, AtomTKD processes a sequence of frames with a ResNet50 backbone followed by a transformer encoder-decoder \citep{zou2021end} to produce spatio-temporal embeddings; these embeddings are then processed with a three-layer perceptron to predict triplets as human-object interaction instances (instrument representing human, target representing object).
This output is weakly supervised based on the ground-truth triplet labels, and at test-time, the bounding box predictions are extracted from the output.

\subsubsection{DualMFFNet}
DualMFFNet makes use of multiplicative feature fusion networks (MFF-Net) \citep{wei2021shallow}, which include a mechanism to aggregate features from different levels of the backbone, to predict each triplet component.
For instrument and target prediction, two separate MFF-Nets with a ResNet50 backbone are trained, yielding instrument and target feature maps.
To model tool-tissue interaction, the instrument and target feature maps are concatenated and processed with a 1D convolution, generating a verb feature map.
Each feature map is passed through a GAP layer and a sigmoid activation, yielding instrument, verb, and target probabilities.
Then, for triplet association, these probabilities are multiplied to generate triplet probabilities, which are then thresholded to obtain final triplet predictions.

For triplet localization, class activation maps are obtained from the instrument feature map using the corresponding instruments of the predicted triplets; the activation map is then thresholded to obtain the instrument bounding box.

\subsubsection{DATUM}
\begin{figure}[!tp]
    \centering
    \includegraphics[width=.9\linewidth]{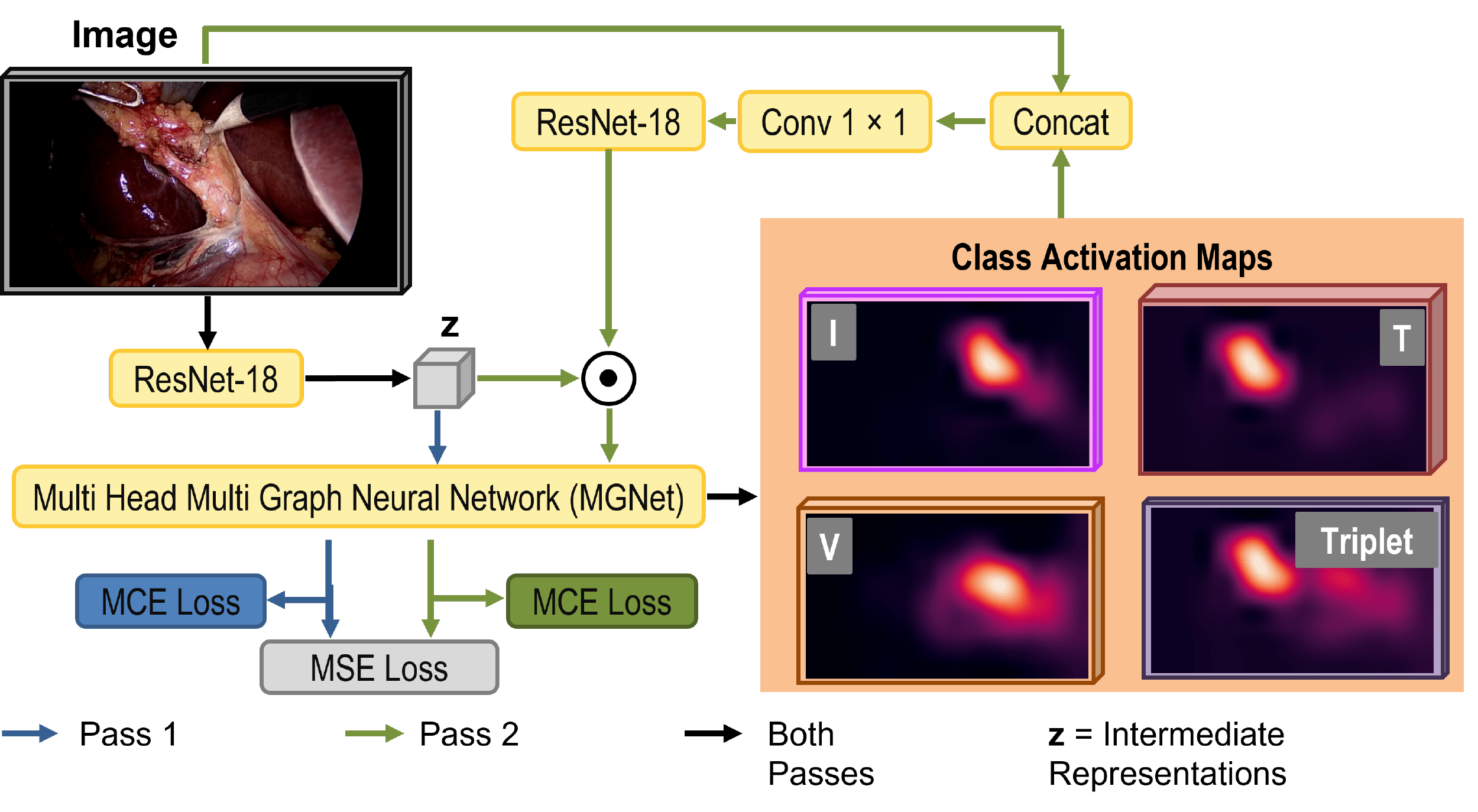} 
    \caption{Architecture \& data flow for the DATUM method featuring graphical modeling, multi-task learning, attention mechanism, class activation maps, and weak supervision.}
    \label{fig:datum}
\end{figure}

DATUM, which stands for Detection of Action Triplets Using Multi-graph networks, relies on a multi-head, multi-graph neural network (MGNet) as shown in Fig. \ref{fig:datum}.
The MGNet is composed of several connected semantic attention modules (SAMs), dynamic graph neural networks (DGCNs) \citep{ye2020attention}, and one classification head per triplet component plus a triplet head; as such, it is inherently multi-task.
Features extracted from a ResNet-18 backbone (trained by self-supervised DINO approach \citep{caron2021emerging}) are fed to the MGNet where they are processed by individual SAMs for instrument, verb, and target. The resulting class activation maps (CAMs) are then forwarded to three DGCNs with edges determined by co-occurrence relations. Features from these three DGCNs are subsequently pooled, concatenated and fed to the triplet DGCN, controlled this time by component-to-triplet mappings. Classification heads are positioned after each of the DGCNs to yield the final class probabilities. 
In this approach, each frame is processed with two passes through the MGNet: a first pass to obtain CAMs, and a second one feeding these maps back into the MGNet to refine the predictions. A mean square error loss ensures consistency between the two passes, in addition to cross-entropy losses for classification.

Triplet localization is obtained during inference from a series of post-processing steps: from the top 3 triplet predictions for a given frame, the corresponding instrument and triplet class activation maps are normalized and binarized using a heuristically determined threshold and then multiplied together. A bounding box is then fit to the resulting blobs.

\subsubsection{Distilled-Swin-YOLO}
\begin{figure}[!tp]
    \centering
    \includegraphics[width=.97\linewidth]{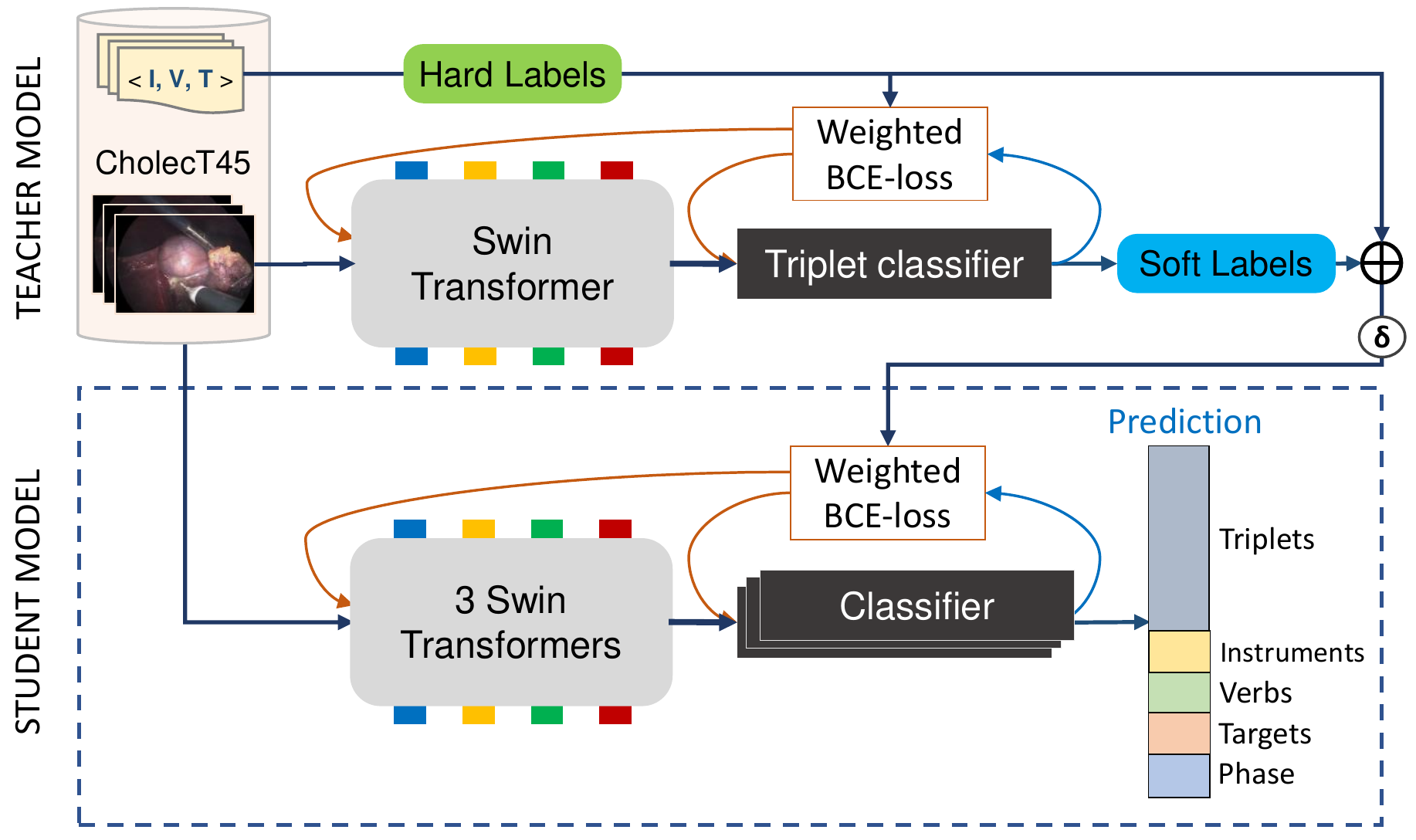} 
    \caption{Architecture \& data flow for the recognition part of Distilled-Swin-YOLO featuring knowledge distillation, transformer, transfer learning, and ensemble modeling.}
    \label{fig:dswin}
\end{figure}

Distilled-Swin-YOLO is a method that makes use of soft-labels for the recognition task and pseudo-labels for the localization task to tackle label uncertainty and overconfidence, which can be especially harmful when using ranking-based metrics like mAP. For triplet recognition, as depicted in Fig. \ref{fig:dswin}, Distilled-Swin-YOLO uses an ensemble of three Swin transformers: a Swin-B model, a Swin-L model, and another Swin-B model trained in a multi-task fashion for instrument, verb, target, and phase recognitions.
To train these models, a Swin-B teacher model is first trained and used to produce softened versions of the original labels, which are in turn used to train each of the 3 final models. 
Distilled-Swin-YOLO additionally uses a weighted binary cross-entropy loss during training to handle class imbalance.

For the triplet localization task, Distilled-Swin-YOLO uses a teacher-student approach to train an instrument detector, the output of which is heuristically combined with the triplet recognition output at test time.
Initially, a YOLOv5 teacher model is pretrained on three external robotic instrument segmentation datasets \citep{allan20202018,allan20192017,sebastianendovis15} for which bounding box annotations are automatically generated from the original segmentation masks.
This teacher model is then used to iteratively generate pseudo-labels on the challenge training set, which are in turn used to train a student model (also a YOLOv5 detector).
For the pseudo-label generation, the teacher predictions are filtered based on the prediction confidence and following the ground truth instrument annotations.

\subsubsection{ EndoSurgTRD} 
 EndoSurgTRD combines two models: a single-frame ResNet50 model for instrument classification and a spatio-temporal TimeSformer model \citep{gberta-2021-ICML} for modeling the verbs and the targets.
The instrument recognition/localization follows the Rendezvous' \citep{nwoye2021rendezvous} WSL approach albeit with a ResNet-50 backbone for feature extraction.
The verb and target components are modeled using a modified TimeSformer which adapts the frame-based ViT \citep{dosovitskiy2020image} to videos such that the inputs are processed as a sequence of patches extracted from the individual frames in the multi-image pathway. This modification extends the ViT's self-attention from its traditional image space to also include time 3D volume.
Each of the triplet's component modules is terminated by a classification head.
Their three output logits are in turn concatenated and passed to a final FC layer that predicts triplet logits.
The complete model is trained end-to-end using the binary cross-entropy loss.

The localization part is achieved with an ad-hoc function that extracts bounding box coordinates for every positive class activation from the ResNet50 model. 
Bounding boxes with low class-wise probability scores and small bounding boxes are ignored in the process. This model, likely due to Docker error, produces the same output box coordinates for all input image, and hence excluded from the localization results.

\subsubsection{InstrumentFirst-Net (IF-Net)}
IF-Net decomposes the problem of action triplet detection into two sequential tasks: (1) instrument detection and (2) triplet classification, taking advantage of the simple constraint that each triplet is bound to a single instrument.
The instrument detection model consists of a ResNet50 model for instrument detection, from which class activation maps are extracted and used to regress instrument bounding boxes following a percentile-based thresholding approach.
In the case of multiple peaks in a single instrument's activation map, the activation map is split into multiple parts and multiple bounding boxes are regressed.

For triplet classification, IF-Net uses a modified ResNet50 model that takes as input the original image concatenated with an instrument class activation map and outputs a probability distribution of triplets.
This probability distribution is then filtered based on the class of the predicted instrument (predicted probabilities for triplets with a different instrument are set to 0).
The triplet classification model is run separately for each instrument detected by the first stage of IF-Net; thus, in the case of no detected instruments, the model is not run.
IF-Net, therefore, models triplet classification as a multi-class problem rather than a multi-class multi-label problem, training the second stage to predict a single triplet.
To obtain the final localized triplet result, the triplet predictions for each detected instrument are aggregated and associated with the previously computed localization results.

\subsubsection{Multi-Task Triplet Transformer (MTTT)}
MTTT is an end-to-end transformer trained in a multi-task fashion.
It consists of a SwinV2-B \citep{liu2022swin} feature extractor followed by 2 masked multi-head attention layers \citep{vaswani2017attention} to incorporate temporal context.
These features are then passed to 4 different prediction branches: instrument detection, verb detection, target detection, and phase detection.
The output logits of each of these branches are then concatenated along with the original features and passed to a triplet prediction branch.
Each branch consists of a fully connected layer and an activation function (sigmoid for instrument, verb, target, triplet; or softmax for phase).

For triplet localization, a CNN is applied to a high-resolution version of the input image.
Following~\cite{nwoye2019weakly}, a modified ResNet-50 \citep{he2016resnet50} (stride in layer 4 is set as one) is employed for feature extraction, followed by two convolutional Long Short-Term Memory (ConvLSTM) \citep{shi2015convolutional} layers for incorporating spatio-temporal context.
Finally, the triplet localization is realized by class activation mapping \citep{shen2016object}.

\subsubsection{ResNet-CAM-YOLOv5}
ResNet-CAM-YOLOv5 combines class activation mapping with supervised pre-training on external datasets to obtain accurate instrument localization, then employs a multi-task approach for triplet recognition.
Specifically, for triplet recognition, a multi-task ResNet50 model is trained on phase recognition, instrument classification, verb classification, and target classification, as in several other methods. In addition, rather than including a module to associate the predicted instrument, verb, and target into a predicted triplet, it includes another branch for direct triplet classification.

For triplet localization, a YOLOv5 instrument detector is first pre-trained on instrument bounding boxes from the CholecSeg8k \citep{hong2020cholecseg8k} and HeiCo \citep{maier2021heidelberg} datasets; then, to obtain a model that localizes instrument tips rather than the entire instrument, the YOLOv5 model is finetuned using the LapChole \citep{stauder2016tum} and Endovis Instrument \citep{sebastianendovis15} datasets.
During inference, to predict the instrument class for each detection, the YOLO-detected bounding boxes are matched with instrument class activation maps ResNet using a threshold-based decision tree, which considers the scores of the recognized instrument classes, the scores of detected bounding boxes, and the mean of the class activation in the region of the detected bounding box.
Meanwhile, for triplet classification, the highest-scoring triplet (obtained from the triplet classification branch) that includes the previously predicted instrument class is predicted as the triplet.

\subsubsection{SurgNet}

\begin{figure}[!tp]
    \centering
    \includegraphics[width=.999998\linewidth]{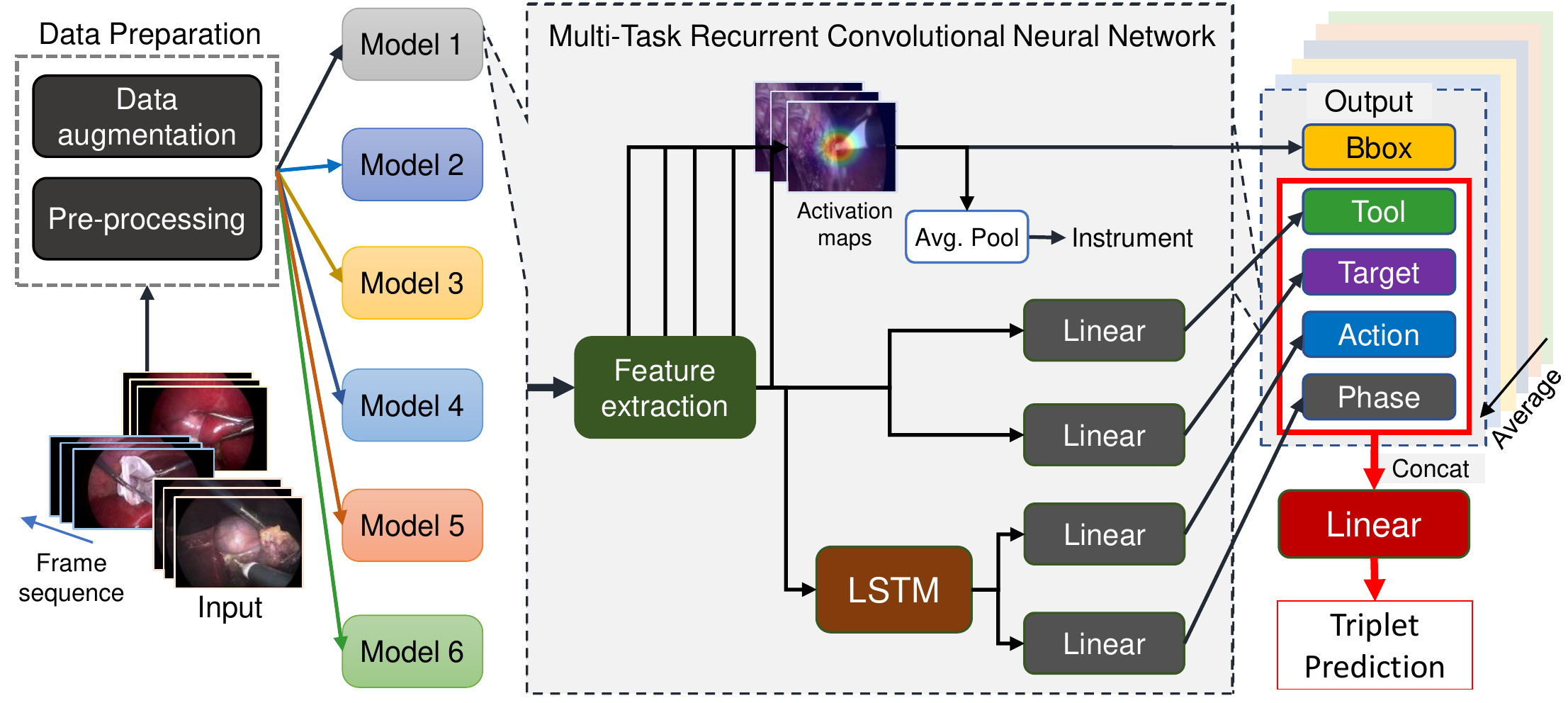} 
    \caption{Architecture \& data flow for the SurgNet method featuring temporal modeling, ensemble modeling, multi-task learning, and weak supervision.}
    \label{fig:surgnet}
\end{figure}

SurgNet is an ensemble of six deep models for surgical action triplet prediction.
As shown in Fig. \ref{fig:surgnet}, each of the models consists of a feature extraction layer, built on one of the popular base architectures ({ResNet50, ResNest50, ResNest101, SENet, EfficientNet-B0, EfficientNetB4}), and a 5-heads multi-task layer for predicting five intermediary tasks ({instrument presence, instrument bounding box location, targets, actions or verbs, phases}), similar to MTTT.
The instrument and target prediction heads are each modeled using a fully connected layer and a sigmoid activation function to enable their multi-label classifications.
The phase and action prediction heads are each modeled using an LSTM unit, followed by an activation function (softmax for multi-class phases and sigmoid for multi-label actions) to compute their class probabilities.
The instrument localization head builds a cascade of intermediary features from various stages of the backbone to capture the instruments' location in the form of heatmaps. The spatial dimension of the heatmaps is scaled to the input image size and their channels are mapped to the number of instrument classes. The output heatmaps are transformed to class presence probability scores using average pooling operation and supervised weakly on instruments' binary presence labels.

Each model in the ensemble is individually trained following a multi-task pipeline. 
An ensemble prediction per intermediary task is obtained by averaging each intermediary prediction across the six ensemble models.
The ensembled instrument presence, targets, verbs, and phases predictions are concatenated to generate a feature vector with a size of 39 and remapped to the 100 triplet class probabilities using a fully connected layer.
Meanwhile, the ensembled instrument bounding box location outputs are thresholded with a value of 0.95, after which the smallest enclosing box can be computed for each instrument.

\subsubsection{URN-Net}
URN-Net uses a ResNet-50 backbone and a Transformer decoder with learnable queries to recognize triplets.
The ResNet50 extracts feature fed as input to the Transformer decoder from which it derives key and value features.
The transformer decoder outputs processed queries that are used to directly predict each triplet component (instrument, verb, target).
In addition, to prevent redundancies in the queries, an additional loss function is imposed on the final processed queries to ensure that they are mutually orthogonal.
Lastly, to encourage temporal consistency, a temporal consistency loss is imposed during training by minimizing the cosine distance between the predicted triplet logits of two frames in a window of 10 frames.

{\rebuttal 
While the URN-Net model developed by team URN successfully passed the validation phase of the challenge, the model only performs surgical action triplet recognition. It is noted that a dummy model was provided for the localization component. Hence, it is omitted from the localization and detection results.
}

\paragraph{\normalfont 
To facilitate cross-referencing and quick access to pertinent technical information, Table~\ref{tab:implementation} consolidates comprehensive implementation details of the presented methods, encompassing architectural specifications, objective loss functions, pre- and post-processing techniques, training strategies, and other relevant particulars
}

\subsection{Theoretical comparative analysis}
We analyze and compare the computer vision technologies in the presented methods under $9$ categories as follows:

\subsubsection{Multi-task learning (MTL)}
To learn relevant triplet component features, multi-task learning of the instruments, verbs, and targets is a common inclusion observed among the presented methods. This helps models to learn refined component-specific features before aggregating them for triplet prediction or to learn triplet features from shared features that are trained to be component-aware. MTL approach was highly popular among contestants: SurgNet, MTTT,  EndoSurgTRD, DATUM, DualMFFNet, ResNet-CAM-YOLOv5, and AtomTKD associate features from triplet components to perform triplet recognition. URN-Net and Distilled-Swin-YOLO, on the other hand, utilize a shared feature extractor where they create separate branches for both triplet and its components. Moreover, due to correlations between surgical phases and actions, SurgNet, MTTT, Distilled-Swin-YOLO, and ResNet-CAM-YOLOv5 incorporate phase recognition along triplet components as an additional task.

\subsubsection{Temporal modeling}
The verb component in a surgical triplet captures the type of action performed by the instrument on the target. Although single frames are sometimes sufficient to identify actions, it is worth exploring how temporal models can help with more ambiguous triplets by using the context from past frames. Two frequent flavors of temporal models are featured in this challenge: Transformers and LSTMs, with variations on the triplet component of interest for the temporal head. SurgNet applies an LSTM only on phase and verb branches to learn temporal features; this choice is motivated by the interplay between phase, verb, and instrument motion patterns. MTTT employs ConvLSTM layers to refine bounding box locations from the noisy instrument class activation map (CAM).  EndoSurgTRD adds a Transformer to model verb and target components separately from single-frame instrument features. Given the importance of past frame features to disambiguate triplet classes, for inference, URN-Net adds a discount factor for every past frame to weight their contribution for triplet recognition.  

\subsubsection{Attention methods}
Attention-based methods are currently one of the state of the art methods on triplet recognition as shown in \citep{nwoye2022cholectriplet2021}. 
These methods learn attention maps that score the importance of a region to achieve high performance on a given task and serve as a powerful tool to explain model predictions. Aside from the RDV-Det, 6 methods utilize attention mechanisms, with the Transformer being the primary model choice. AtomTKD and URN-Net apply the Transformer for the recognition of triplets and their components whereas MTTT and Distilled-Swin-YOLO additionally use the Transformer to learn phase features. Moreover, the choice of where to place the Transformer varies across methods and depends on the type of triplet component, as  EndoSurgTRD applies the Transformer on only the verb and target components. As a non-Transformer attention method, DATUM applies a semantic attention module to extract relevant regions of interest from scene features. Interestingly, for localization, none of the methods uses an attention mechanism.

\subsubsection{Knowledge distillation (KD)}
Distillation is the process of carrying over knowledge between a pair of models, with the source of knowledge commonly referred to as the "teacher" and the recipient as the "student". While oftentimes distillation serves as a model compression technique by having a lightweight model copy a heavier one, its purpose in the two entries utilizing it herein was to implement a form of soft label learning. Atom-TKD performed distillation between two models: T-AtomNet as the teacher, focusing on separate triplet components, and S-TRNet as the student, with more focus on the triplet as a whole. Distilled-Swin-YOLO used self-distillation: three models were involved in this process, one for each triplet component, each one acting both as the teacher and the student. In both cases, KD is performed through logits, via an MSE (Atom-TKD) or cross-entropy (Distilled-Swin-YOLO) loss, to handle uncertainty in predictions.
KD is also used by Distilled-Swin-YOLO to train a YOLOv5 for instrument localization.

\subsubsection{Activation modeling}
Class activation map (CAM) is an effective way of obtaining the localization of scene objects using only image-level labels. CAMs highlight the discriminative regions in the image that can contribute to object localization. Furthermore, \cite{nwoye2020recognition, nwoye2021rendezvous} demonstrate a useful property of CAMs: high activation regions correspond to the area around the instrument tip. This style of localization is helpful when instrument bounding box annotations are time-consuming and expensive. RDV-Det, SurgNet, IF-Net,  EndoSurgTRD, and DualMFFNet integrate weakly supervised methods based on CAMs for localizing the instruments present in the surgical scene. Surgical triplets usually feature instrument movements over a certain time range, which MTTT exploits by utilizing ConvLSTM models on the CAM output to modulate features across time. Moreso, DualMFFNet combines instrument and triplet CAMs multiplicatively to locate triplets.  

\subsubsection{Ensemble methods}
Methods that ensemble model predictions provide a comprehensive way to factor knowledge from multiple models, and limit noise in predictions. SurgNet and Distilled-Swin-YOLO employ ensemble with variations on the style of architecture to model triplet components. SurgNet creates an ensemble of large models such as ResNet50 and ResNest50/10 as well as parameter efficient models EfficientNetB0/B4 and SENet and averages the predictions for both the triplet recognition and localization tasks. Distilled-Swin-YOLO, on the other hand, utilizes both base and large variants of Swin-Transformer to train solely on triplet and an additional base variant of Swin-Transformer to train in a multi-task setting, involving triplet components.

\subsubsection{Transfer learning}
Transfer Learning is a de facto approach to utilize models trained on one task such as object detection for another related task such as instrument/triplet detection. Distilled-Swin-YOLO and ResNet-CAM-YOLOv5 train YOLOv5, a state-of-the-art object detector, on external laparoscopic datasets as well as on the challenge training data for triplet localization. This allows the model to exploit YOLOv5's capabilities to better detect instruments. 
A rudimentary form of transfer learning, known as pre-training, involves prior training of a model on another dataset, usually a larger one, to mitigate overfitting. Rather than actual model training, pre-training is most approached indirectly via full or partial weights initialization; especially for the backbone parameters. 
72.7\% models presented in this challenge are pretrained on ImageNet; Cholec80 and other endoscopic datasets make up the rest. Only one model is without pretraining. The specific pretraining source for each presented model is provided in Table \ref{tab:implementation}.

\subsubsection{End-to-end vs separate training strategy}
Models employed for complex tasks such as the ones proposed in this challenge tend to include several parts, each part performing a different function. For such models, choosing a training sequence can be an issue: end-to-end training, with all parts trained simultaneously, or separate training. Note that in many situations, end-to-end training can be experimentally inconvenient, if not completely infeasible, especially with long videos such as those offered by CholecT50. Three entries have opted for end-to-end learning: MTTT (CNN + ConvLSTM - Vision Transformer + attention), DualMFFNet (2 CNN), and URN-Net (CNN + Transformer). MTTT in particular is a spatio-temporal combination, limiting the temporal range for end-to-end learning: in this case, clips of 5 frames were used.

\subsubsection{Levels of model supervision}
Alternatives to full supervision are highly sought after in surgical data science due to the heavy cost of manually labeled surgical data. In that spirit, no labels at all were provided for the localization task of this challenge, which left no possibility for true full supervision of challenge entrants. To solve this, weak supervision based on Class Activation Maps (CAM) was an overwhelmingly popular choice. The activation in  CAMs manifests as a blob covering the entity to detect; further manipulations of the CAM, e.g. fitting a bounding box to the blob, enable spatial detection at no annotation cost since CAMs are a byproduct of training for binary presence detection. 
We observed that 8 methods used this type of CAM-based weak supervision: AtomTKD, SurgNet, IF-Net, MTTT,  EndoSurgTRD, DATUM, DualMFFNet, and RDV-Det. Two methods, however, opted for more conventional object detectors: Distilled-Swim-YOLO and ResNet-CAM-YOLOv5 both rely on YOLOv5, prefatorily trained using full spatial supervision from external datasets.

\paragraph{\normalfont
In summary, it is worth highlighting that the teams are not constrained in terms of deep learning techniques, model parameters, pre/post-processing, or hyperparameter tuning in their modeling. Therefore, this methodological analysis may be taken more as an insight rather than facts. And since the presented models are also constrained by the time frame of the contest, we state that further improvement is potentially possible on each methodology given extended training time and resources.}
\begin{table*}[!thb]
    \caption{Implementation details across entrants.} 
    \label{tab:implementation}
    \resizebox{1.0\textwidth}{!}{%
    \includegraphics[width=1.05\linewidth, height=24.5cm]{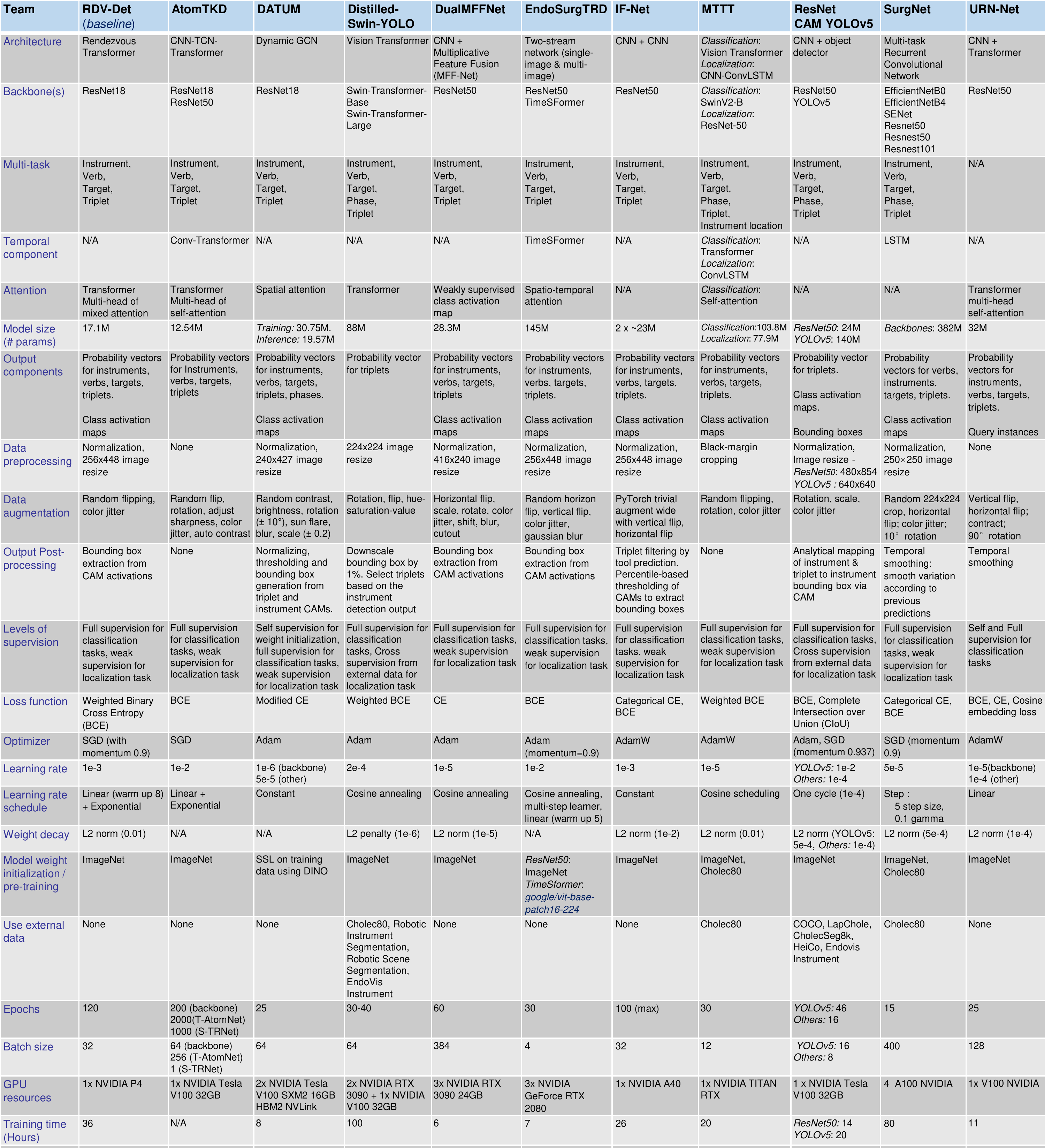}
    }
\end{table*}

\section{Experiment and evaluation setup}

\subsection{Implementation details}
\hint{Similar to last edition (transposed) but will include details on model training like an epoch, LR, batch norm, etc., likely to be in Appendix if too big}
The implementation details for the reproducibility of all submissions are summarized in Table \ref{tab:implementation}.

\subsection{Evaluation protocol}
All presented models are trained on the publicly released part of the CholecT50 dataset otherwise known as CholecT45. Evaluations are conducted on the hidden 5 video test set.
At the evaluation, submitted Docker containers are executed on the challenge test set producing a JSON file per video containing the outputs for the three sub-tasks.

Triplet recognition evaluation is based on the triplet probability scores ($Y_{IVT}$) for valid 94 out of 100 triplet classes ignoring the last 6 labels with at least one NULL component in the triplet composition, as done in the previous challenge \citep{nwoye2022cholectriplet2021}.
Model's direct outputs for any component of the triplet are not considered, instead, the predictions for instruments, verbs, and targets are filtered from $Y_{IVT}$ using \cite{nwoye2021rendezvous}'s disentanglement function for uniformity across all architectural designs.
Detection evaluation for instruments and triplets is based on the predicted bounding box coordinates and class identities.
For each metric, we use video-specific averaging to obtain the mean score: 
per-category metric scores are averaged across videos, before computing the mean across categories. Averaging ignores unrepresented categories.
All the performance scores are computed using \path{ivtmetrics}~\citep{nwoye2022data}\footnote{\url{https://pypi.org/project/ivtmetrics}},
which is a dedicated metrics library for surgical action triplet evaluation.

\subsection{Evaluation metrics} \label{sec:metrics}

The performance of the challenge models is evaluated using the average precision (AP) metric. This metric assesses the precision (p) and recall (r) scores, which are calculated based on the true positive (TP), false positive (FP), and false negative (FN) counts as follows:

\begin{equation}
\label{metrics:pr}
    p = \frac{TP}{TP+FP}~, \qquad 
    r = \frac{TP}{TP+FN}
\end{equation}

The AP metric is used for both the recognition and detection task categories albeit computed differently.
In the recognition task category, TP, FP, and FN are determined by applying various threshold values to the predicted probability scores, and the AP is calculated as the area under the precision-recall curve. 
Specifically, we measure the AP for recognizing the triplet individual components, denoted as AP$_I$ (instrument), AP$_V$ (verb), and AP$_T$ (target), as well as the AP for triplet associations: instrument-verb (AP$_{IV}$), instrument-target (AP$_{IT}$), and instrument-verb-target (AP$_{IVT}$) combinations, following the established protocol \citep{nwoye2020recognition}. The AP$_{IVT}$ metric serves as the primary evaluation metric for the complete instrument-verb-target combination.

In the detection task category, the TP, FP, and FN are computed by measuring the level of overlap between the groundtruth and predicted bounding boxes. 
A detection is assigned a TP if the degree of overlap exceeds a certain threshold and the ID (i.e. instrument ID for category 2 or triplet ID for category 3) is correct.
A detection without a corresponding groundtruth is assigned as FP while a missed detection in the presence of a ground truth is marked as FN. Note that a non-substantial overlap is also treated as a missed detection and recorded as FN.
The corresponding AP is computed as a weighted mean of the precisions achieved at each threshold, with the increase in recall from the previous threshold used as the weight:
\begin{equation}
    \label{metrics:AP}
    AP = \int_{0}^{1} p(r)dr .
\end{equation}

\begin{table}[tp]
    \centering
    \caption{TAS: Triplet association metrics}
    \label{tab:tas}
    \setlength{\tabcolsep}{3pt}
    \resizebox{\columnwidth}{!}{%
    \begin{tabular}{@{}lll@{}}
        \toprule
        Metric &  Name & Description \\  
        \midrule
        LM & Localize \& Match & \makecell[tl]{Percentage of tools localized at a given\\ IoU and matched with correct triplet IDs.} \\
        $p$LM & Partial Localize \& Match & \makecell[tl]{Percentage of tools localized below the a given\\ IoU but matched with correct triplet IDs.} \\
        IDS & Identity Switch & \makecell[tl]{Percentage of tools localized at a given\\ IoU but IDs are swapped within the frame.} \\
        IDM & Identity Missed & \makecell[tl]{Percentage of tools localized at a given\\ IoU but IDs are missed entirely.} \\
        MIL & Missed Localization & \makecell[tl]{Percentage of correctly recognized\\ triplet IDs without localization.} \\
        RFP & Remaining False Positives & \makecell[tl]{Percentage of false alarms\\ after other TAS metrics have been considered.} \\
        RFN & Remaining False Negatives  & \makecell[tl]{Percentage of missed predictions\\ after other TAS metrics have been considered.} \\
        \bottomrule         
    \end{tabular}
    }
\end{table}
From an alternative viewpoint, the average recall (AR) metric is also used for this task to measure the percentage of TP that are correctly identified by the model at different levels of minimum IoU thresholds. Different from the AP metrics, recall score is computed instead of the precision score.

We also utilize the triplet association scores (TAS) introduced in \cite{nwoye2022data} to assess the capacity of the models at associating the bounding boxes to the correct triplet IDs. The TAS metric, presented in Table \ref{tab:tas}, offers an in-depth analysis of the task 3 category by evaluating the detection from different perspectives.

Additionally, we analyze the quality of model predictions using the top$K$ accuracy ($K\in [5,10,15,20]$) and average over top$K$@[5:20] at intervals of 5 steps as done in prior research \cite{nwoye2022cholectriplet2021}.

Overall, these evaluation metrics provide a comprehensive assessment of the models' performance in the challenge, enabling a thorough analysis of their recognition and detection capabilities as well as in the ranking stability.

\begin{table*}[t]
    \centering
    \caption{Performance summary of the presented methods for the three tasks (Localization IoU threshold $\theta=0.5$).}
    \label{tab:results:quantitative:summary}
    \setlength{\tabcolsep}{10pt}
    \resizebox{0.92\textwidth}{!}{%
    \begin{tabular}{@{}lccccrccrcc@{}}
    \toprule
        & \multicolumn{4}{c}{Triplet recognition} & \phantom{abc} & \multicolumn{2}{c}{Instrument localization } & \phantom{abc} & \multicolumn{2}{c}{Triplet detection } \\  
        \cmidrule{2-5} \cmidrule{7-8} \cmidrule{10-11} 
        Model & $AP_{I}$ & $AP_{V}$ & $AP_{T}$ & $AP_{IVT}$ && $AP_{I}$ & $AR_{I}$ && $AP_{IVT}$ & $AR_{IVT}$ \\ 
        \midrule        
        ResNet-CAM-YOLOv5 &   80.3 & 50.3 & 38.4 & 29.0 && \textbf{41.9} & \textbf{49.3} && \textbf{4.49} & \textbf{7.87} \\
        
        Distilled-Swin-YOLO &   83.8 & 52.0 & \textbf{45.9} & \textbf{35.0}  && \underline{17.3} & \underline{30.4} && \underline{2.74} & \underline{6.16} \\
        
        MTTT &   \underline{84.0} & 49.5 & \underline{40.3} & \underline{34.5}  && 11.0 & 21.1 &&  1.47 & 3.65 \\
        
        DualMFFNet  & 64.9 & 40.4 & 29.9 & 20.0  && 04.6 & 06.6 && 0.36 & 0.73 \\
        
        RDV-Det $\ddag$ & 78.2 & 46.6 & 35.9 & 29.0 && 03.0 & 07.6 && 0.24 & 0.79 \\
        
        IF-Net &   72.0 & 42.3 & 30.4 & 23.6 && 00.7 & 03.6 && 0.22 & 0.92 \\
        
        AtomTKD  & \textbf{85.7} & \underline{52.3} & 39.2 & 27.2 && 00.9 & 02.4 && 0.15 & 0.32 \\
        
        SurgNet   & 78.8 & \textbf{55.7} & \underline{40.3} & 34.4  && 10.8 & 19.6 && 0.13 & 0.39 \\
        
        DATUM &  66.2 & 39.7 & 32.3 & 23.1 && 00.3 & 03.2 && 0.08 & 0.48 \\
        
        URN-Net   & 82.4 & 49.3 & 38.1 & 27.3 && - - - & - - - &  &- - -& - - - \\
         EndoSurgTRD  & 73.6 & 42.1 & 26.3 & 18.8 && - - - & - - -  & & - - - & - - - \\
        \midrule
        Average $\pm$ standard deviation (stdev) & $77.3\pm7.2$ & $47.3\pm5.4$ & $36.1\pm5.8$ & $27.4\pm5.7$ && $10.1\pm12.5$ & $16.0\pm15.0$ && $1.1\pm1.4$ & $2.4\pm2.7$ \\
        \bottomrule 
        \multicolumn{9}{l}{\makecell[l]{\footnotesize{ {\bf bold} = best score. \underline{underlined} = second best. $\ddag$ organizers' baselines. }}}\\
    \end{tabular}
    }
\end{table*}
\section{Results and discussion}

\subsection{Triplet recognition}

We present a brief overview of the average precision (AP) scores for both triplet component recognition and component association in Table \ref{tab:results:quantitative:summary}. 
As anticipated, the instrument is the most accurately recognized component with a minimum AP score of $64\%$ across all participating teams. Remarkably, over $80\%$ of the teams achieves scores exceeding $70\%$, and half of them scores above $80\%$.
For the verb component, performance ranges between ${\sim}42-52\%$ AP with a mean of $47.3\pm 5.4\%$. As with the previous edition of the challenge \citep{nwoye2022cholectriplet2021}, accurate recognition of the target being acted upon remains an open problem. Here, we see a relatively wide range of performances between ${\sim}26-46\%$ with a majority of methods recognizing the target with an AP of $<40\%$. However, we also do note that there appears to be a strong correlation between recognition of the overall triplet and recognition of the target (R=0.93). Further, the fact that the triplet recognition performance here appears to be more strongly linked with the target performance than the instrument (R=0.74) or verb (R=0.84) performance may be indicative of a bottleneck that target recognition brings and should inform future research. 
Finally, we note that the top 3 methods on the triplet recognition task also attempt to learn phase features, potentially demonstrating how coarser-grained information could inform fine-grained tasks such as triplet recognition. 
Operative phases are defined through the different steps/actions surgeons perform to complete a procedure. For each phase, defined by a specific procedural task, there is a characteristic set of triplets used.
The correlation could ensures that the likelihood of detecting a specific triplet will increase in associated phases and thus, conditions the model on a subset of probable triplets rather than on a whole using phase features.
Additionally, leveraging phase information could also help in discriminating closely related triplets with imperceptible change in roles which are usually tough to distinguish visually, e.g. \triplet{grasper, retract, gallbladder} occurring at the \textit{calot triangle dissection} phase and \triplet{grasper, pack, gallbladder} during \textit{gallbladder packaging}.

In terms of association for the different components, we see a comparable performance in associating the instrument with each of the two other components. Overall triplet recognition performance maxes out at $35.0 \%$ by the Distilled-Swin-YOLO model. 

\begin{figure}[h]
    \centering
        \includegraphics[width=0.95\linewidth]{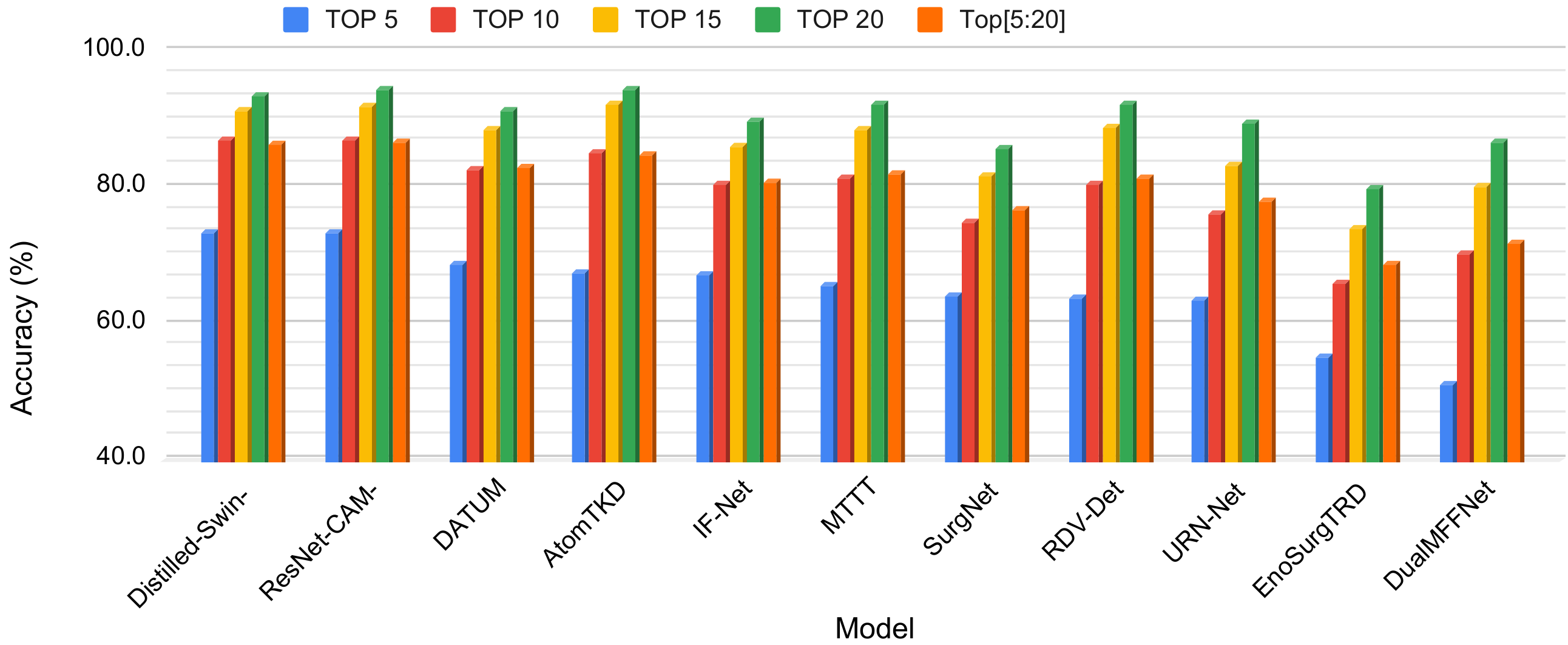} 
        \caption{Action triplet recognition top-K accuracy.}
    \label{fig:topk}
\end{figure}

Further, using Top K accuracy, in Fig. \ref{fig:topk}, we also measure the ability of models to predict triplets within their K highest-scoring outputs. This allows us to consider the difficulty of the problem introduced by the high semantic overlap between certain classes (eg. sharing multiple components). Unsurprisingly, we see that the Distilled-Swin-YOLO model also comes out on top with top 5 and top 10 accuracies of 73.5 and 87.9, respectively. The gains in the performance are for two reasons - first, the choice of model, a Swin-Transformer that generates meaningful hierarchical features, and second, the use of an ensemble that suppresses noise in the predictions. Interestingly, the ResNet-CAM-YOLOv5 model, which places fourth for triplet recognition comes in joint-highest in the top 5 and top \{5:20\} performances, respectively.
The higher topK accuracy compared to the AP scores show that a reduced number of triplet classes might theoretically be easier for a deep learning model to train. This would imply super-classifying related classes, resulting in more data to train each class. It would also increase inter-class variability and help to balance the loss optimization strategy affected by intra-class conflicts. However, a larger number of triplet categories increases the granularity and becomes more beneficial and clinically relevant.

\begin{figure}[h]
    \centering
        \includegraphics[width=\linewidth]{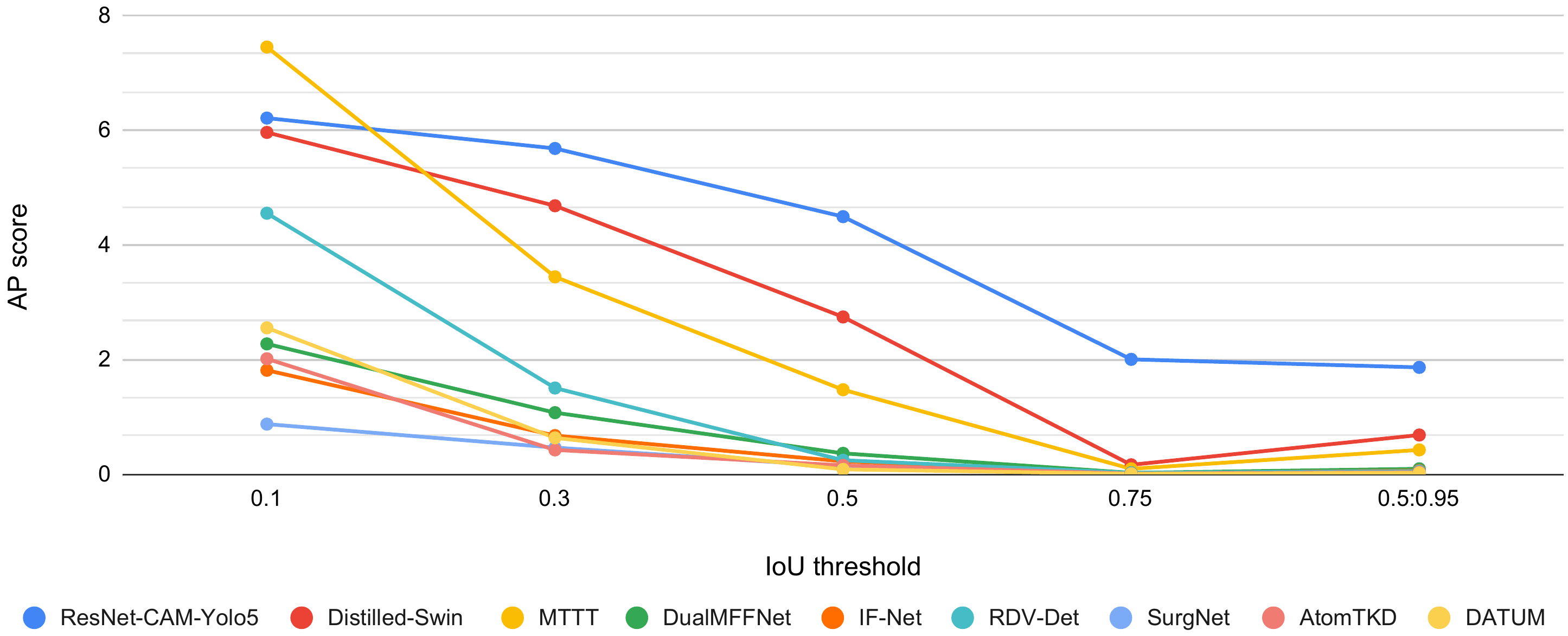} 
        \caption{Surgical instrument localization mAP.}
    \label{fig:loc-map}
\end{figure}

\subsection{Instrument localization}
Fig.~\ref{fig:loc-map} shows the instrument localization performance for each competing method, evaluated using the AP metric at various IoU thresholds.
Immediately, we observe a large contrast in performance between the top-performing ResNet-CAM-YOLOv5 and the rest of the models; this can be attributed to the fact that the ResNet-CAM-YOLOv5 includes components that are pretrained on the CholecSeg8k, HeiCo, M2CAI, and EndoVisSub datasets for instrument localization, in contrast to the other methods, which rely purely on weak supervision signals.
This is further substantiated when studying the AP$_{0.1}$ results, where a couple of other methods (MTTT, SurgNet), are competitive with the ResNet-CAM-YOLOv5; this indicates that these models are able to achieve a rough localization of the instruments, but lack precision and are therefore harshly penalized by increasing the IoU threshold for the AP computation.

\begin{figure}[h]
    \centering
        \includegraphics[width=\columnwidth]{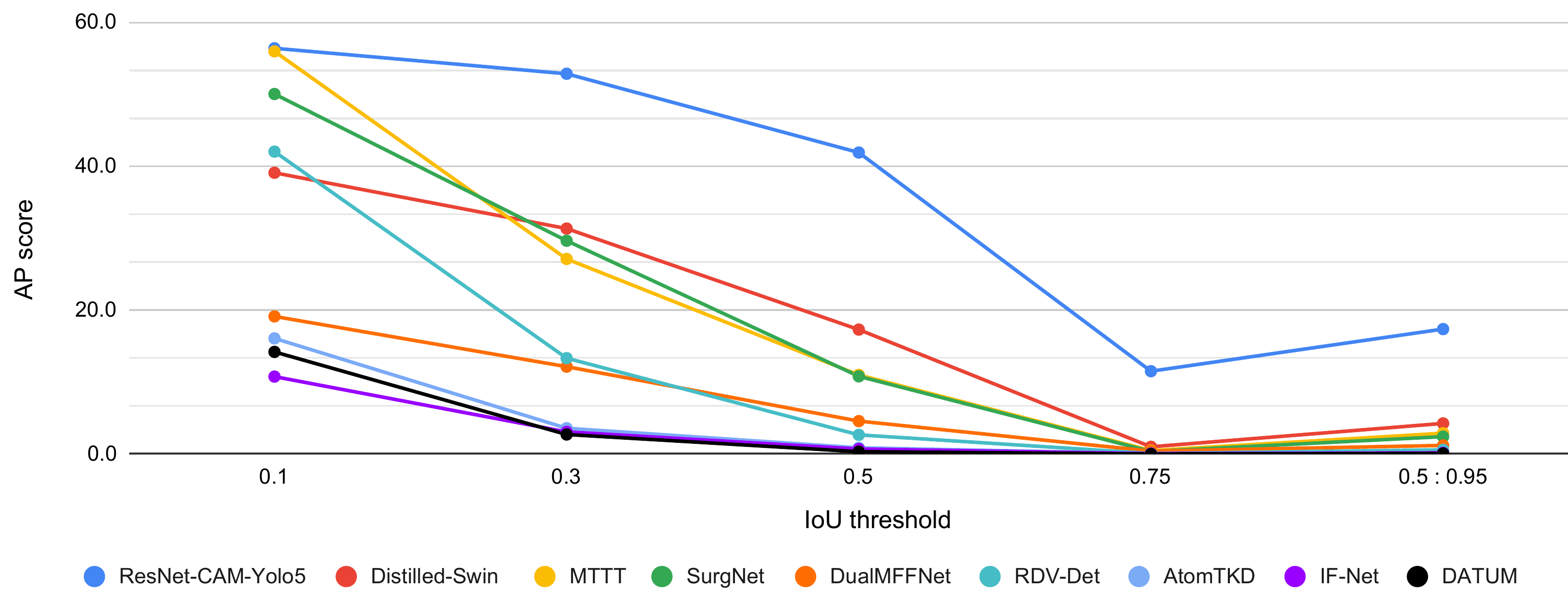} 
        \caption{Surgical action triplet detection mAP.}
    \label{fig:det-map}
\end{figure}

\subsection{Triplet detection (box-triplet matching)}
The triplet detection assessment evaluates joint performance on triplet presence recognition and instrument localization. It is also affected by the correct association of the triplet IDs to their corresponding instrument bounding boxes. 
The detection AP is presented in Fig.~\ref{fig:det-map}.
Across different IoU, the models' ranking is stable, however, the performances decrease with a stricter overlapping threshold.
It is observed that correct instrument localization plays a crucial role in this metric for the top 3 models (ResNet-CAM-YOLOv5, Distilled-Swin-YOLO, MITT) when compared with Fig.~\ref{fig:loc-map}, 
though without discounting the associating part which could account for the altering of the model ranking on this metric after the top 3.

\begin{table}[h]
    \centering
    \caption{Bbox-triplet association performance ($\theta=0.5$)}
    \label{tab:results:quantitative:assoc}
    \setlength{\tabcolsep}{6pt}
    \resizebox{\columnwidth}{!}{%
    \begin{tabular}{@{}lrrcccrr@{}}
        \toprule
        Model &  LM $\uparrow$ & $p$LM $\downarrow$ & IDS $\downarrow$ & IDM $\downarrow$ & MIL $\downarrow$ & RFP $\downarrow$ & RFN $\downarrow$ \\  
        \midrule
        ResNet-CAM-YOLOv5 & \textbf{23.9} & \textbf{8.2} & 0.9 & 0.1 & \textbf{0.1} & \textbf{3.3} & 63.5 \\
        Distilled-Swin-YOLO & \underline{12.0} & 11.3 & 0.3 & \textbf{0.0} & \underline{0.3} & 33.0 & \underline{43.1} \\
        MTTT & 8.6 & 25.4 & 0.1 & 0.1 & 0.4 & 11.6 & 53.8 \\
        RDV-Det $\ddag$ & 3.3 & 29.0 & \textbf{0.0} & \textbf{0.0} & 1.1 & 5.4 & 61.1 \\
        DualMFFNet & 3.0 & 17.4 & 0.1 & \textbf{0.0} & 1.2 & \underline{3.5} & 74.8 \\
        IF-Net & 2.0 & 14.9 & 0.1 & 0.7 & 1.5 & 35.9 & 45.1 \\
        SurgNet & 2.0 & 8.5 & \textbf{0.0} & \textbf{0.0} & 1.6 & 18.9 & 69.0 \\
        DATUM & 0.4 & 22.8 & \textbf{0.0} & \textbf{0.0} & 1.3 & 45.1 & \textbf{30.5} \\
        AtomTKD & 0.1 & 14.9 & \textbf{0.0} & \textbf{0.0} & 3.1 & 28.4 & 53.6 \\
        \bottomrule 
        \multicolumn{8}{l}{\makecell[l]{\scriptsize{{\bf bold} = best score. \underline{underlined} = second best. $\ddag$ organizers' baselines.}}}
    \end{tabular}
    }
\end{table}
We, hence, quantify the association performance at varying localization conditions as presented in Table \ref{tab:results:quantitative:assoc} using the triplet association score (TAS) metrics described in Section \ref{sec:metrics} at a threshold of 0.5.
We observe that while the association scores when keeping only the well-localized bounding boxes (LM) are low, they become higher when considering also the partially localized boxes (pLM).

The TAS metrics also show that identity switches (IDS) and identity misses (IDM) are minimal for all the methods. This argument is further strengthened by low MIL scores showing that the number of correctly recognized triplets that are not localized at any overlapping degree is negligible. Hence, the bottleneck lies in obtaining substantial localization overlap by a weakly-supervised method. 
The remaining FP is relatively low for 3 models (ResNet-CAM-YOLOv5, RDV-Det, DualMFFNet) whereas the remaining FN is high for all the methods. Both false negatives and false positives account for ${\sim} 65\%$ of the triplet association strength of the models. The FN/FP cases are frequently observed among semantically similar triplets such as \triplet{grasper, grasp, gallbladder} in place of \triplet{grasper, retract, gallbladder}, \triplet{scissors, cut, cystic-artery} in place of \triplet{scissors, cut, blood-vessel}. Discriminating such triplets could be a potential area of improvement for future works.

Despite the observed low performances on the spatial detection tasks, the analysis of the presented models still reveals a consistent pattern between the average precision (AP) and average recall (AR) scores, as shown in Table \ref{tab:results:quantitative:summary}. This indicates a correlation between precision and recall in the model predictions, suggesting the potential for improvement in capturing relevant information from the input data. The observed consistency in AP and AR scores across the models implies a stable performance that can serve as a foundation for further optimization and enhancement in future iterations.

\begin{figure*}[!htbp]
    \centering
        \includegraphics[width=\textwidth]{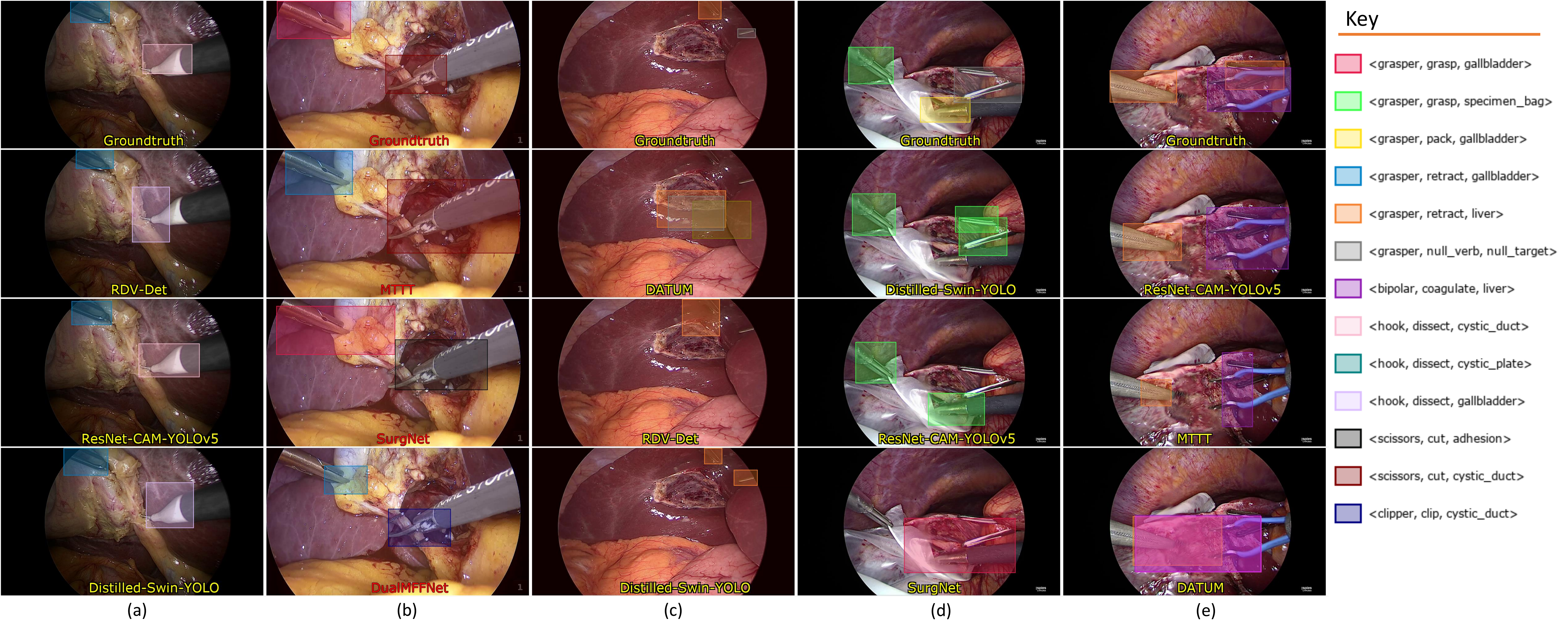} 
        \caption{Qualitative results visualizing triplet detection under different instrument usage conditions, grouped by columns: {\normalfont (a) regular case occurring too often, (b) rare case occurring only once in a procedure, (c) nearly invisible instrument, (d) crowded scene, and (e) partial occlusion. Ground truths are shown at the top rows while predictions from three randomly selected models are shown in other rows. Localization is illustrated using bounding boxes with color coding (Key at the right) matching the groundtruth and predicted triplet class labels.}}
    \label{fig:qualitative:usage-pattern}
\end{figure*}

\begin{figure*}[!thb]
    \centering
        \includegraphics[width=\textwidth]{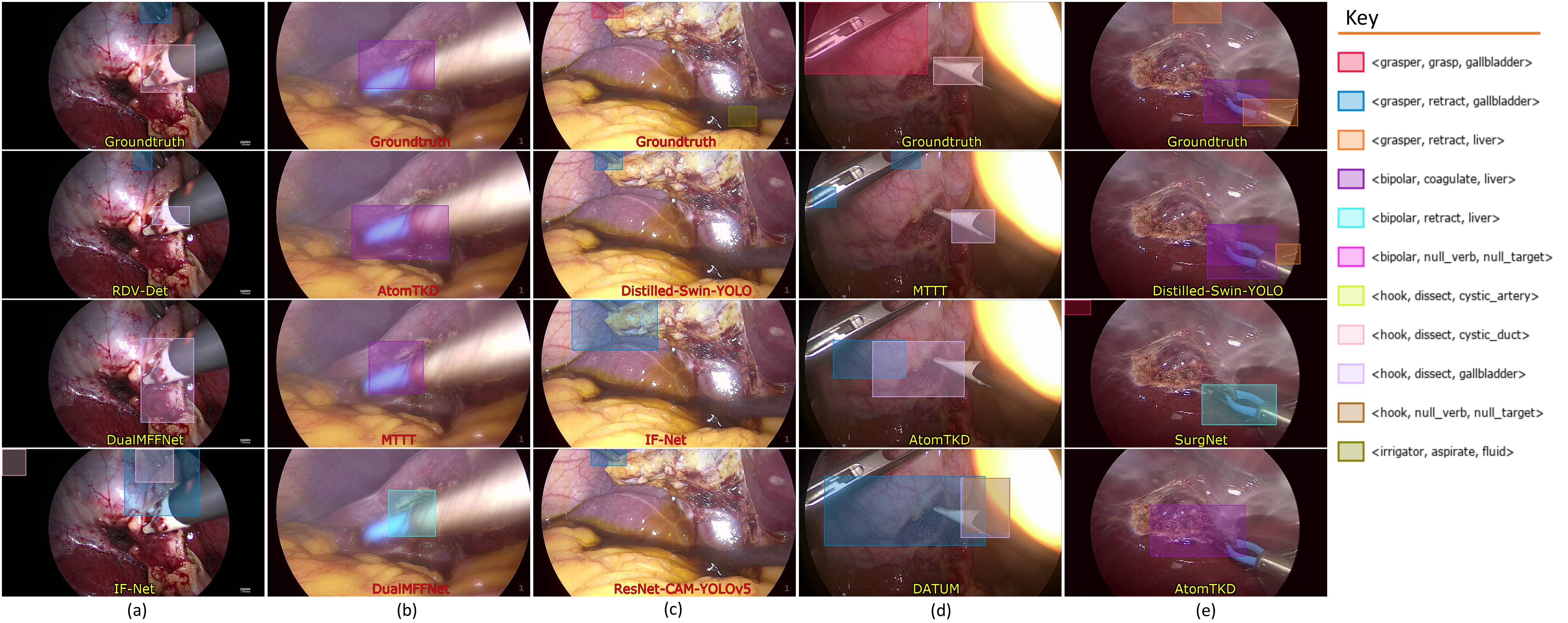} 
        \caption{Qualitative results visualizing triplet detection under different visual challenges (surgical noise), grouped by columns: {\normalfont (a) blood stain, (b) rapid motion blurring, (c) immersion in fluid, (d) specular light reflection, and (e) presence of smoke. Ground truths are shown at the top rows while predictions from three randomly selected models are shown in other rows. Localization is illustrated using bounding boxes with color coding (Key at the right) matching the observed triplet classes.}}  
    \label{fig:qualitative:noise}
\end{figure*}

\subsection{Qualitative results}
We also analyze a rich set of qualitative results from the models on different surgical situations, vis-à-vis, instrument usage patterns (Fig. \ref{fig:qualitative:usage-pattern}) and visual challenges (Fig. \ref{fig:qualitative:noise}).
In both figures, varying surgical conditions are presented in each column, the groundtruths are on the first rows whereas predictions from 3 randomly selected models are on the other rows.
Instrument localization is represented with bounding boxes of different colors for different triplet classes.

It is observed that more common triplets (Fig. \ref{fig:qualitative:usage-pattern}a) in a clear scene are better detected than rarer triplets (Fig. \ref{fig:qualitative:usage-pattern}b) that might occur only once and during a short moment in a procedure.
Predicting the correct category is oftentimes an issue with rare triplets. 
Another observation is that a miss is usually the case when only a tiny portion of the instrument is visible (Fig. \ref{fig:qualitative:usage-pattern}c), notwithstanding, the Distilled-Swin-YOLO model appears robust to this case.
A crowded scene (Fig. \ref{fig:qualitative:usage-pattern}d) presents a combination of missed prediction (e.g. ResNet-CAM-YOLOv5) and overlapping localization of multiple instruments as one (e.g. SurgNet).
Even with correct localization, the triplet identities are oftentimes switched or mismatched (e.g. Distilled-Swin-YOLO). 
This could become an issue in certain surgical procedures, e.g. gastric bypass, requiring up to 6 trocar ports or when more than 2 instruments are visible in each frame, such as during retraction for suturing or knot-tying, stitching, etc.
Occlusion occurs when instruments seem to overlap in a 2D view, the instrument closest to the camera (partially) obstructs the view of the instrument at the back. Under this situation (Fig. \ref{fig:qualitative:usage-pattern}e), the occluded instrument is either localized as part of the occluding instrument (e.g. MTTT, ResNet-CAM-YOLOv5) or missed entirely (e.g. DATUM).

We observe quite interesting behaviors of the models under challenging visual conditions (Fig. \ref{fig:qualitative:noise}). 
Bleeding in surgical procedures can cause blood stains on the instruments or camera lens.
Under this condition (Fig. \ref{fig:qualitative:noise}a), model localizing capacity is challenged: most models would localize just the non-stained part of the instruments (e.g. IF-Net, RDV-Det).
The view (or image quality) can also be impaired by rapid motion of camera and instruments, or lack of focus adjustment (Fig. \ref{fig:qualitative:noise}b). 
This leads to imprecise localization which could be attributed to the blurred instrument boundaries (e.g. DualMFFNet).
In the cleaning and coagulation phase, instruments, such as the irrigator, are immersed in fluid/water (Fig. \ref{fig:qualitative:noise}c).
We observed that this leads to high false negatives for all the considered models.
Specular reflection from endoscopic light can change the appearance of the surgical scene as seen in Fig. \ref{fig:qualitative:noise}d.
These light reflections outshine the surgical scene and thus affects the models in both the localization of obvious instruments like grasper and hook (e.g. AtomTKD, DATUM, MTTT) and their triplet classification.
Localizing the same instrument with multiple bounding boxes is also observed (e.g. MTTT).
Sometimes, when there is smoke generated through use of coagulation instruments, the effect ranges from minimal to adverse depending on the thickness of the smoke. 
In Fig. \ref{fig:qualitative:noise}e, it can be observed that the grasper held by the assistant surgeon is missed by all models as it is heavily overlaid by smoke.
The light smoke on bipolar has minimal effects on the models though it affects the instrument localization for the AtomTKD and the triplet classification for the SurgNet.

In general, the instruments are mostly predicted correctly. Most of the missed predictions arise from incorrect verb and/or target predictions.
Notwithstanding, the major factor contributing to the low AP scores is insufficient overlap between the localized instruments and their corresponding groundtruths leading to prohibitively high false negatives as shown in the last column of Table \ref{tab:results:quantitative:assoc}.
This is also confirmed in Fig. \ref{fig:det-map} where the AP performance decreases with increasing the overlapping threshold.
Triplet misclassification and their identity mismatch contribute to further decrease of the detection AP score.

\subsection{Statistical Analysis}
\begin{table}[t]
    \centering
    \caption{Results of Wilcoxon signed-rank test for CholecTriplet2022 challenge task using the baseline RDV-Det method as the alternative method, showing $p$-values and effect sizes of each comparison.}
    \label{tab:results:quantitative:wilcoxon-rec}
    \setlength{\tabcolsep}{10pt}
    \resizebox{\columnwidth}{!}{%
    \begin{tabular}{@{}llll@{}}
    \toprule

    {\vtop{\hbox{\strut Proposed}\hbox{\strut Methods}}} &
    {\vtop{\hbox{\strut Triplet}\hbox{\strut recognition}}} &
    {\vtop{\hbox{\strut Instrument}\hbox{\strut localization}}} &
    {\vtop{\hbox{\strut Triplet}\hbox{\strut detection}}} \\
    \toprule
     AtomTKD & $p>0.05$ & $p<0.001$ & $p<0.001$ \\
    DATUM & $p<0.05$ & $p<0.001$ & $p<0.001$ \\
    Distilled-Swin-YOLO & $p<0.01$ & $p<0.001$ & $p<0.001$ \\
    DualMFFNet & $p>0.05$ & $p<0.01$ & $p>0.05$ \\
    EndoSurgTRD & $p\approx0.01$ & $p<0.001$ & $p<0.001$ \\
    IF-Net & $p>0.05$ & $p>0.05$ & $p>0.05$ \\
    MTTT & $p>0.05$ & $p<0.001$ & $p<0.001$ \\
    ResNet-CAM-YOLOv5 & $p>0.05$ & $p<0.001$ & $p<0.001$ \\
    SurgNet & $p\approx0.001$ & $p<0.01$ & $p>0.05$ \\
    URN-Net & $p>0.05$  & - - - - - - -  & - - - - - - - \\
    \bottomrule  
    \end{tabular}
    }
\end{table}

We evaluate the rank stability of participating teams' methods using the Wilcoxon signed-rank test. This non-parametric test is appropriate for our data since it is not multivariate normal and teams' predictions may contain many outliers. The test compares each team's method against a null hypothesis ($H_0$) to determine the statistical significance of its performance compared to the alternative method. Here, we use the baseline (RDV-Det) as the alternative method. Each $H_0$ test produces a p-value between 0 and 1, with a smaller $p$-value indicating stronger evidence to reject the null hypothesis. Basically, a $p$-value $<$ 0.05 is considered statistically significant, with which the $H_0$ can be rejected at 5\% confidence level. We perform the test on 35 random batches of N=128 sample data from the test set for both the recognition and detection tasks.

We report our analysis at different confidence levels (5\%, 1\%, 0.1\%) as presented in Table 8. It shows that at each of these confidence intervals, a number of methods that were closely ranked with the baseline have statistically insignificant differences in performance, while the rest of the methods have significantly higher or lower performance than the baseline. We also observed a number of $p$-values $<$ 0.001, which are extremely statistically significant. This analysis reveals that the CholecTriplet 2022 challenge attracted teams with both similar and diverse methods, resulting in a range of performance levels.

\subsection{Comparison with the CholecTriplet2021 challenge}

\begin{figure*}[!th]
    \centering
        \includegraphics[width=.8\textwidth]{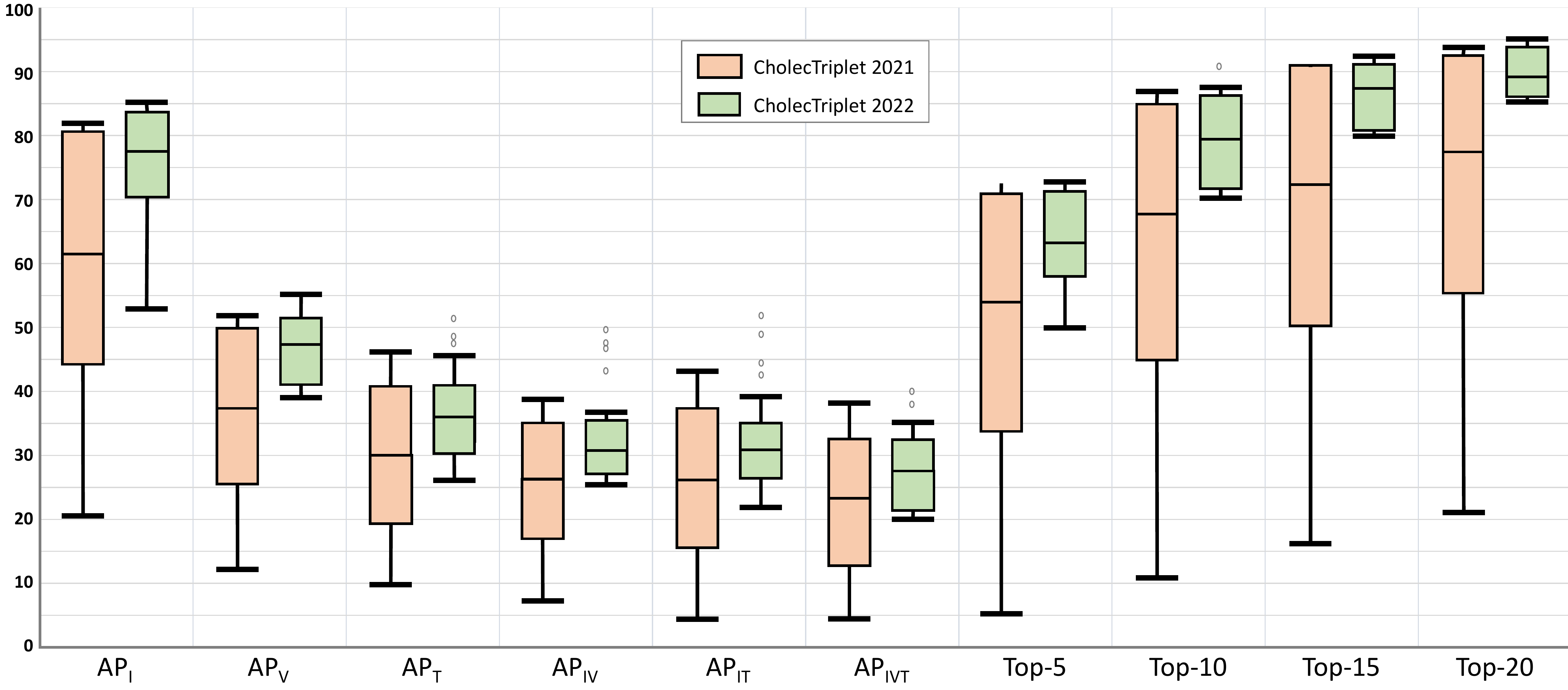} 
        \caption{Box plot comparison of the quantitative results (Category 1) with the previous edition of the challenge (CholecTriplet 2021).}  
    \label{fig:boxplot}
\end{figure*}

While the CholecTriplet2022 challenge is an evolution of the 2021 edition \citep{nwoye2022cholectriplet2021}, we note here several differentiating factors in the task, organization \& participation, methodology, knowledge reuse, and outcome. We also discuss possible reuse of knowledge from the previous edition.

\subsubsection{Task}
The primary distinction between the two editions is the scope of each challenge. The first edition aimed at benchmarking, analyzing, and improving methods for surgical triplet recognition. As a natural progression, the current edition extends these tasks to be able to localize the operating instrument corresponding to each detected triplet (i.e. from triplet recognition to triplet detection) without spatial labels for training. To evaluate these methods, the 5 test videos of the CholecT50 dataset were augmented to include bounding box labels around the instrument tips in addition to the binary triplet labels available for all 50 videos.

\subsubsection{Organization and participation}
In this edition, several aspects of the validation and submission process were refined. Regarding validation, the CholecTriplet2022 challenge utilized a self-validation approach with strict requirements on input/output formats and computational resources. Compared to the CholecTriplet2021 challenge, this enabled a more transparent validation process, and faster debugging, and helped focus organizational communication toward only significant issues. The submission portal was also moved to DockerHub to allow native submission of docker files without conversion or compression. The evaluation protocol itself was modified to allow structured outputs as a JavaScript Object Notation (JSON) file rather than a text file for quicker processing. The metrics were also appropriately modified to be able to effectively assess the localization of instruments and triplets in addition to the recognition metrics used in the previous edition.
To support these additional organizational efforts, the committee was expanded from 4 members to 8. The participation however dropped by 45\% likely due to the reduced perceived novelty of the dataset compared to its first introduction in 2021 edition.

\subsubsection{Methodology}
The second edition of the CholecTriplet challenge saw significant methodological advancements, mainly aimed at accommodating the tool localization aspect of the task. The recognition component, which remained the same in both editions, had a stable performance, reflecting the continuity in methodology. Among the notable changes, the use of Transformer-based models for feature extraction increased substantially from 8.7\% in 2021 to 63.6\% in 2022, while the use of CNN backbones reduced from 95.6\% to 90.5\%. Attention modeling and the visualization/description of learned features by attention maps were increasingly adopted in the 2022 edition, with 72.7\% of the teams exploring attention modeling, compared to only 30.4\% in 2021. Furthermore, temporal modeling for action triplet recognition was less emphasized in the 2022 edition, with only 4/11 teams considering it, compared to 12/23 teams in 2021. Multi-task learning remained the go-to methodology for both editions, with 20/23 teams in 2021 and 10/11 teams in 2022 using it. Graphical modeling remained underutilized, with less than 10\% in both editions. The 2022 edition saw the introduction of new deep learning approaches, such as knowledge distillation and weakly-supervised localization.

\subsubsection{Knowledge reuse}
\begin{table}[t]
    \centering
    \caption{Compariosn of team performances between 2021 and 2022 edition of CholecTriplet challenge showing effects of knowledge reuse.}    \label{tab:results:quantitative:knowledge-reuse}
    \setlength{\tabcolsep}{5pt}
    \resizebox{\columnwidth}{!}{%
    \begin{tabular}{@{}lllccr@{}}
    \toprule
    Team & Model & Changes & 2021 $AP_{IVT}$ & 2022 $AP_{IVT}$ & $\Delta~AP_{IVT}$ \\
    \toprule
    CITI & MTTT & None & 32.0 & 34.5 & +2.5 \\ 
    SK & DualMFFFNet & None & 18.4 & 20.0 & +1.6 \\
    2AI & SurgNet &  {\vtop{\hbox{\strut Additional MTL branch}\hbox{\strut for localization}}} & 36.9 & 34.4 & -2.5 \\
    CAMMA & RDV $/$ RDV-det & {\vtop{\hbox{\strut Localization algorithm,}\hbox{\strut different pretraining}}} & 32.7 & 29.0 & -3.8 \\
    CAMP & EndoVisNet $/$ IF-Net & Entirely different & 9.3 & 23.6 & +14.3 \\    
    \bottomrule  
    \end{tabular}
    }    
\end{table}

In this study, we aimed to assess the extent of knowledge reuse in the context of surgical action triplet recognition from the previous edition of the CholecTriplet challenge. Our analysis included the performance of five teams that participated in both editions, focusing on their reuse of previous models and any subsequent improvements in performance. 

The results presented in Table \ref{tab:results:quantitative:knowledge-reuse} indicate that the first two teams (CITTI and SK) were able to improve their previous performances by +2.5\% and +1.6\%, respectively, by reusing their previous models (MTTT and DualMFFNet, respectively). MTTT builds a standalone temporal model while DualMMFNet relies on the CAM from its instrument feature map to independently model the localization part of the task. In contrast, the performance of Team 2AI, who integrated a new localization module into their previous model architecture via a MTL branch, dropped by -2.5\%. This suggests that a more complex optimization of both tasks can harm the performance of the one, in this case, the recognition task. Similarly, we observed that the CAMMA team's extended model (RDV-Det) with a new detection component showed a -3.8\% drop in the recognition task, possibly due to the new model's pretraining on ImageNet alone, as opposed to Cholec80 + ImageNet in their previous edition. Finally, the CAMP team proposed an entirely new model that significantly improved the performance by +14.3\% AP, surpassing their previous edition's model.

Overall, this analysis highlights the potential benefits of knowledge reuse from previous editions and the possible trade-offs involved in integrating new modules into existing architectures.

\subsubsection{Outcome}
In terms of the outcome, the 2022 edition offers a wide range of methods: 6 out of 11 models have extremely distinctive methodologies. Interestingly, methods focusing on knowledge distillation are proposed to supplement weakly-supervised learning of spatial localization from different datasets.
There is also a significant reduction in the gap between the top and bottom placed models in the leaderboard as illustrated by box plots in Fig~\ref{fig:boxplot}: the smaller standard deviation across all metrics illustrates an improved and better understanding of the task modeling compared to the 2021 edition. 
The final triplet recognition mAP is not outperformed in the latest edition of the challenge. This could be attributed to the increased complexity of the task and the challenge of optimizing more task heads at various levels of supervision (full and weak). The top-K results, meanwhile, generally improved.

\subsection{Limitations}
Notwithstanding the milestone, the challenge on surgical action triplet detection faces some limitations.
Firstly, as with most challenges, given that the participants are not constrained in terms of modeling, the comparative analysis may be taken as indications rather than facts, especially since aside from the strength of the modeling techniques, lots of other factors such as data preprocessing, augmentation styles, model size, hyperparameter tuning, training computational power, weight initialization, etc., can influence model performances.

Another limitation is the choice of supervision. The localization task of the challenge is defined by weak supervision by default. 
However, weak supervision is approached differently: in this case, some teams train directly on weaker binary presence labels while others train a teacher model on external data which in turn generates pseudo-spatial labels for model finetuning.
Both approaches are acceptable standards for weak supervision, however, their advantages differ thus affecting their comparability.
The best we could do in terms of constraints is to withhold the spatial labels of the benchmark dataset.

One of the limitations of the validation phase lies in its evaluation of predefined conditions, focusing on aspects like run-time error, input/output pipeline, resource limit, and access privileges, without considering the potential presence of a dummy model specifically designed to bypass these checks.

\begin{figure*}[!th]
    \centering
        \includegraphics[width=.95\textwidth]{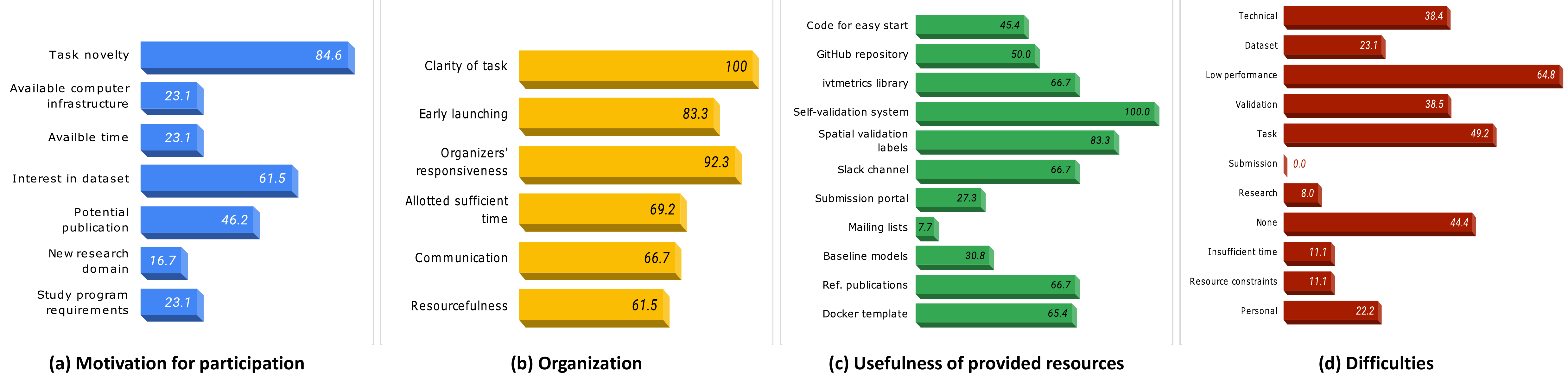} 
        \caption{Summary of a survey conducted on CholecTriplet2022 at the end of the  challenge. \normalfont{Each item is independently rated on a scale of 1-5. Average values expressed in percentage. Similar items are grouped in one plot for analysis. Values of grouped items are not normalized to a fixed number}.}  
    \label{fig:survey}
\end{figure*}

On the evaluation, how the traditional AP metrics translates a model capacity at triplet recognition and detection is yet to be investigated in full. In the CholecT45 and CholecT50 datasets, there is a level of similarity between certain distinct triplets. As an example, the triplets \triplet{grasper, grasp, gallbladder} and \triplet{grasper, retract, gallbladder} are different but at certain stage of the procedure could be used interchangeably without misrepresentation. At some other point, an action, e.g. {\it dissect}, can happen at an unclear boundary such that the two adjacent anatomies, e.g. {\it gallbladder-neck} versus {\it cystic-duct}, can be interchangeably considered as the target without loss of generality. Taking everything into consideration, it may be fairer to design AP metrics that would consider the proximity of the predicted triplets to their interchangeable ground truth counterparts. This would to an extent address the low performance scores of deep learning models on triplet recognition and detection.

Looking at the obtained results, there is still room for improvement, especially on the localization task.
Also, localization, which is explored on the instruments in this challenge, is yet to be extended to the surgical targets.
The task of weakly supervised target anatomy localization remains an open problem as generating spatial annotation is quite expensive, and importantly, the target anatomy is not solely determined by its visibility, which further complicates the problem.
Furthermore, there are fewer datasets in the community for differentiating the anatomy/organ structures causing a delay in research focusing on recognition of anatomical targets: there is also no known pre-training source. This contributes to the imbalance and suggests why the target recognition is poor compared to the instruments.

Finally, the inference speed of the presented models was not calculated because they include additional Docker initialization time that cannot be separated from actual model inference, making fair comparison of execution time challenging. Nevertheless, this is a necessary trade-off to the Docker packaging of a model in a consistent environment with all its OS and library dependencies as a single container for the challenge evaluation.
\section{Challenge survey}
We conducted an anonymized survey on the organization of the challenge. The questionnaires were circulated online to the participants including those unable to proceed to the submission stage. The respondents were allowed to rate on a scale of 1-5 (later converted to percentile in the analysis) their satisfaction on each of the parameters of the challenge, and offer recommendations where they feel so.
The survey received responses from 13 teams, representing a response rate of 43\% compared to the total number of registered teams (38). We collated the responses into four groups as shown in Fig. \ref{fig:survey} for their analysis. 
The key findings can be summarized as follows:

\begin{enumerate}[(a)]
    \item {\it Motivation for participation:} 
    The analysis of participants' motivation to engage in the triplet detection challenge, reveals that the novelty of the task and the interest in the CholecT50 dataset were the primary drivers of participation. The opportunity for a joint publication of challenge findings held secondary importance. Factors such as the availability of compute infrastructure, study program requirements, time, and the exploration of a new research domain had minimal influence on participants' motivation. 
    
    \item {\it Challenge organization:} 
    The survey analysis examines participants' feedback on the organization and coordination of the challenge. Participants highly rated the clarity of the task instructions and the responsiveness of the organizers to their queries. The challenge was deemed to have started in a timely manner, allowing participants sufficient time for completion. While the provision of helpful resources and communication received slightly lower ratings, they still scores above 60\%. These findings suggest that participants appreciated the clear instructions and prompt support from the organizers.
    
    \item {\it Usefulness of provided resources:} 
    On the perceived usefulness of the resources provided in the challenge, the survey analysis shows that the participants found the self-validation system and spatial-labeled dataset to be the most useful resources provided. The provision of a metrics library, reference publications, Docker template \& guide, and a Slack communication channel were also highly rated. Participants also found the snippets of code for easy starting and GitHub repositories to be helpful. Reference baseline models and the automated submission portal received a slightly lower rating. The creation of group and support mailing lists was rated lowest, potentially due to the preference for using a dedicated Slack channel for daily communication.
    
    \item {\it Difficulties:} 
    The survey analysis reveals that participants identified the low performance of their models as a major deterrent to progressing beyond the validation phase, likely due to the inherent difficulty of the task itself. Technical difficulties primarily centered around building Docker containers, while the submission process posed the fewest challenges, likely due to the effectiveness of the self-validation system. Other factors such as resource and time constraints, research aspects, dataset usage, and the validation process were reported to have less impact. Notably, no issue received a rating of about 45\%, suggesting that overall, participants encountered moderate to manageable difficulties throughout the challenge. 
\end{enumerate}
 
The survey analysis provides valuable insights for improving future computer vision and biomedical competitions. It highlights the perceived usefulness of provided resources, guiding organizers in optimizing resource allocation. Additionally, the analysis identifies the specific challenges faced by participants, aiding in the identification and mitigation of obstacles in future iterations of similar tasks. These findings offer valuable feedback to enhance participant engagement and inform the organization of research challenges.
\section{Conclusion}
With CholecTriplet2022, we presented a study in the form of an international contest focusing on the development of deep learning models for the detection of fine-grained surgical activities in laparoscopic videos.
The task is subdivided into three categories of (1) recognizing surgical actions as triplets \triplet{instrument, verb, target}, (2) localizing the instruments performing the actions, and (3) pairing the localized instruments with their correct action triplets. 
The research and experiments are conducted on the CholecT50 dataset which has been annotated with triplet labels. 
The localization sub-task is handled by weak supervision.
The challenge ran through a 5-month window recording a total of 11 teams' participation showcasing 11 new deep learning methods to solve the problems. 

The presented models employed several computer vision techniques such as state-of-the-art 
CNN architectures, multi-task learning, temporal modeling, graphical modeling, knowledge distillation, attention/transformer, transfer learning, model ensembling, end-to-end training, weak supervision, etc.
While performances close to the baseline are recorded on the localization task, higher and promising results are obtained on the triplet recognition task.
Our in-depth analysis of the results, quantitatively and qualitatively, reveals the behaviors of presented methods under various surgical conditions and visual challenges prevalent in surgical video data. With this meta-analysis, we provide insights for the advancement of the present and similar studies in the field.

In summary, this study contributes, in no small amount, to the gradual advancement of research on tool-tissue interaction understanding, which would be helpful for the development of future generation operating rooms that include AI assistance. 
Following a similar setup, the design and experiments contained in this paper can be extended to other surgical procedures.
\section*{CRediT authorship contribution statement}
{\small
{\bf C.I. Nwoye}: Conceptualization, Data Curation, Methodology, Software, Investigation, Validation, Formal Analysis, Visualization, Data Analysis \& Interpretation, Writing - Original Draft, Writing - Review \& Editing, Resources, Challenge Organization.
{\bf T. Yu}: Conceptualization, Data Curation, Formal Analysis, Writing - Original Draft, Writing - Review \& Editing, Resources, Challenge Organization.
{\bf S. Sharma}: Conceptualization, Data Curation, Software, Formal Analysis, Validation, Visualization, Writing - Original Draft, Resources, Challenge Organization.
{\bf A. Murali}: Conceptualization, Data Curation, Formal Analysis, Software, Writing - Original Draft, Resources, Challenge Organization.
{\bf D. Alapatt}: Conceptualization, Data Curation, Formal Analysis, Writing - Original Draft, Challenge Organization.
{\bf A. Vardazaryan}: Conceptualization, Data Curation, Investigation, Writing - Review \& Editing, Resources, Challenge Organization.
{\bf K. Yuan}: Formal Analysis, Visualization, Writing - Original Draft, Challenge Organization.
{\bf J. Hajek, W. Reiter, A. Yamlahi, F. Smidt, X. Zou, G. Zheng, B. Oliveira, H, Torres, S. Kondo, S. Kasai, F. Holm, E. Özsoy, S. Gui, H. Li, S. Raviteja, R. Sathish, P. Poudel, B. Bhattarai, Z. Wang, G. Rui}: Methodology, Software, Investigation, Visualization, Writing - Review \& Editing.
{\bf M. Schellenberg, J. Vilaça, T. Czempiel, Z. Wang, D. Sheet, S.K. Thapa, M. Berniker, P. Godau, P. Morais, S. Regmi, T. Tran, J. Fonseca, J.H. Nölke, E. Lima, E. Vazquez, L. Maier-Hein, N. Navab}: Team Supervision, Writing - Review \& Editing.
{\bf B. Seeliger, C. Gonzalez, P. Mascagni} : Data Curation, Writing - Review \& Editing.
{\bf D. Mutter} : Data Curation, Writing - Review \& Editing, Resources, Supervision.
{\bf N. Padoy}: Conceptualization, Project Administration, Project Supervision, Funding Acquisition, Resources, Writing - Review \& Editing, Challenge Organization.
}

\section*{Declaration of competing interest}
{\small
The authors declare that they have no known competing financial interests or personal relationships that could influence the work reported in this paper.
}

\section*{Data availability}
{\small
The CholecT50 dataset and the validation data used in the challenge have been made available to the public and accessible via \url{https://github.com/CAMMA-public/cholect50}. 
The test set spatial labels will be released publicly.
The baseline model code will be released as well. Participants can release their code on their own volition.
All released code would be linked to the central GitHub repository for the challenge: \url{https://github.com/CAMMA-public/cholectriplet2022}.
}
\section*{Acknowledgment}
{\small
The organizers would like to thank the IHU and IRCAD research teams for their help with the initial data annotation during the CONDOR project. 
We also thank Stefanie Speidel, Lena Maier-Hein, Danail Stoyanov, and the entire EndoVis 2022 organizing committee for providing the platform for this challenge.
}

\section*{Funding}
{\small
This work was supported by French state funds managed within the Plan Investissements d’Avenir by the ANR under references: National AI Chair AI4ORSafety [ANR-20-CHIA-0029-01], Labex CAMI [ANR-11-LABX-0004], DeepSurg [ANR-16-CE33-0009], IHU Strasbourg [ANR-10-IAHU-02] and by BPI France under references: project CONDOR, project 5G-OR [DOS0180017/00]. 

Software validation and evaluation were performed with servers managed by CAMMA at University of Strasbourg and IHU Strasbourg, as well as HPC resources from Unistra Mésocentre, and GENCI-IDRIS [Grant 2021-AD011011638R1, 2021-AD011011638R2, 2021-AD011011638R3].

Awards for the challenge winners were sponsored by IHU Strasbourg, NVIDIA, and Medtronic Ltd.

Participating teams would like to acknowledge the following funding:
\textbf{CITI}: Shanghai Municipal Science and Technology  Commission [20511105205].
\textbf{SDS-HD}: Twinning Grant [DKFZ+RBCT]; the Surgical Oncology Program of the National Center for Tumor Diseases (NCT) Heidelberg, by the German Federal Ministry of Health under the reference number 2520DAT0P1 as part of the pAItient project, and by HELMHOLTZ IMAGING, a platform of the Helmholtz Information \& Data Science Incubator.
European Research Council (ERC) under the European Union’s Horizon 2020 research and innovation programme (NEURAL SPICING; grant agreement No. [101002198]) and Surgical Oncology Program of the National Center for Tumor Diseases (NCT) Heidelberg.
\textbf{2AI-ICVS}: Fundação para a Ciência e a Tecnologia (FCT), Portugal and the European Social Fund, European Union, for funding support through the “Programa Operacional Capital Humano” (POCH) in the scope of the Ph.D. grants [SFRH/BD/136721/2018, SFRH/BD/136670/2018]. Grants [NORTE-01-0145-FEDER-000045, NORTE-01-0145-FEDER-000059], supported by Northern Portugal Regional Operational Programme (NORTE 2020), under the Portugal 2020 Partnership Agreement, through the European Regional Development Fund (FEDER). Also funded by national funds, through the FCT and FCT/MCTES in the scope of the project [UIDB/05549/2020, UIDP/05549/2020].
\textbf{SHUANGCHUN}: Guangdong Climbing Plan under Grant [pdjh2023c21602].
\textbf{CAMP}: partially supported by Carl Zeiss AG.
}

\bibliographystyle{model2-names.bst}\biboptions{authoryear}
\bibliography{main}

\end{document}